\documentclass[english]{article}
\usepackage[T1]{fontenc}
\usepackage[latin9]{inputenc}
\usepackage{color}
\usepackage{mathrsfs}
\usepackage{amsmath}
\usepackage{graphicx}
\usepackage{subfigure,array}
\usepackage[a4paper]{geometry}
\makeatletter

\providecommand{\tabularnewline}{\\}

\makeatother

\usepackage{babel}
\begin{document}

\title{Higher order nonclassicalities of finite dimensional coherent states: A comparative
study}

\author{Nasir Alam{$^{\dagger}$}, Amit Verma{$^{\mathsection}$}
and Anirban Pathak{$^{\dagger,}$}{\footnote{anirban.pathak@jiit.ac.in}}\\
{$^{\dagger}$} Jaypee Institute of Information Technology,
A-10, Sector-62, Noida, UP-201307, India\\
$^{\mathsection}$ Jaypee Institute of Information Technology,
Sector-128, Noida, UP-201304, India}

\maketitle

\begin{abstract}
Conventional coherent states (CSs) are defined in various ways. For
example, CS is defined as an infinite Poissonian expansion in Fock
states, as displaced vacuum state, or as an eigenket of annihilation
operator. In the infinite dimensional Hilbert space, these definitions
are equivalent. However, these definitions are not equivalent for the
finite dimensional systems. In this work, we present a comparative
description of the lower- and higher-order nonclassical properties
of the finite dimensional CSs which are also referred to as qudit
CSs (QCSs). For the comparison, nonclassical properties of two types
of QCSs are used: (i) nonlinear QCS produced by applying a truncated
displacement operator on the vacuum and (ii) linear QCS produced by
the Poissonian expansion in Fock states of the CS truncated at $(d-1)$-photon
Fock state. The comparison is performed using a set of nonclassicality
witnesses (e.g., higher order antiubunching, higher order sub-Poissonian
statistics, higher order squeezing, Agarwal-Tara parameter, Klyshko's
criterion) and a set of quantitative measures of nonclassicality
(e.g., negativity potential, concurrence potential and anticlassicality). The higher order nonclassicality witness have found to reveal the existence of higher order nonclassical properties of QCS for the first time. 
\end{abstract}

\section{INTRODUCTION}
Coherent states drew considerable attention of the quantum optics
and atom optics community for various reasons. For example, a CS is
known to be a quasi-classical state or the most classical state among
the quantum states \cite{Agarwal2013quantum}, and it has applications
in almost all fields of physics \cite{klauder1985coherent,zhang1990coherent}.
In quantum optics, CS has been traditionally defined in various ways,
such as displacement of vacuum state, eigenket of annihilation operator,
or infinite Poissonian superposition of Fock states \cite{Agarwal2013quantum,ficek2014quantum}.
In the infinite dimensional Hilbert space, these different definitions
of CS are equivalent. However, in the finite dimensional Hilbert space,
different definitions lead to different finite dimensional coherent
states which are referred to as qudit coherent states \cite{miranowicz2014phase}.
In general, a qudit may be viewed as a $d$-dimensional quantum state
that can be expanded in Fock-state $\left(\left|n\right\rangle \right)$
basis as 

\begin{equation}
|\psi\rangle_{d}=\sum_{n=0}^{d-1}c_{n}|n\rangle.\label{eq:state1}
\end{equation}

With the recent developments in quantum state engineering \cite{miranowicz2014state,sperling2014quantum,vogel1993quantum,miranowicz2004dissipation}
and quantum computing and communication {[} see Ref. \cite{pathak2013elements}
and references therein{]}, production and manipulation of these types
of quantum states have become very important. Further, in the recent
past, several applications of nonclassicality \cite{pathak2013elements,hillery2000quantum,alam2017lower}
and a few experimental demonstrations of higher order nonclassicality
\cite{avenhaus2010accessing,allevi2012measuring,allevi2012high,pevrina2017higher} have been reported.
Specifically, in the Laser Interferometer Gravitational-Wave Observatory
(LIGO), squeezed vacuum state has been successfully used for the detection
of the gravitational waves \cite{abbott2016observation,abbott2016gw151226}
by reducing the noise \cite{harry2010advanced,grote2013first}. Squeezed
state is also used in continuous variable quantum cryptography \cite{hillery2000quantum},
teleportation of coherent state \cite{furusawa1998unconditional},
etc. Anti-bunching is used for characterizing single photon sources
\cite{yuan2002electrically} which are essential for the realization
of various schemes for secure quantum communication. Further, entangled
states have been established to be useful for various quantum information
processing tasks (\cite{pathak2013elements} and references therein).
For example, entangled states are essential for quantum teleportation
\cite{bennett1993teleporting}, densecoding \cite{bennett1992communication},
quantum cryptography \cite{ekert1991quantum}, etc.
In addition to the nonclassical states having the above mentioned
applications, QCSs (being a finite superposition of Fock states, which
are always nonclassical) have also drawn considerable attention of
the quantum optics community in its own merit. In fact, various aspects
related to the properties (mostly nonclassical properties), possibilities
of generation and potential applications of finite dimensional optical
states have been carefully investigated in last three decades \cite{miranowicz1994coherent,leon1997finite,miranowicz2014phase,buvzek1992coherent,galetti1996discrete,kuang1993dynamics,zhu1994even,roy1998coherent,miranowicz2003quantum1,miranowicz2003quantum2,pegg1998optical}
with specific attention to the QCSs \cite{miranowicz1994coherent,leon1997finite,miranowicz2014phase,buvzek1992coherent,roy1998coherent}.
Specifically, generation possibilities of finite dimensional states
of light have been discussed in Refs. \cite{leon1997finite,miranowicz2014phase,miranowicz2003quantum2,pegg1998optical}
and their properties have been studied in Refs. \cite{buvzek1992coherent,miranowicz2014phase,zhu1994even,roy1998coherent,miranowicz2003quantum1}.
To obtain a QCS in particular or a finite dimensional Fock superposition
state in general, we would require a mechanism to truncate the infinite
dimensional conventional Fock-state expansion of a driving field.
In this context, a set of closely connected and extremely interesting
concepts have been developed. Such concepts include quantum scissors
\cite{ozdemir2001quantum,miranowicz2014phase,miranowicz2004dissipation,leonski2011quantum},
which aims to truncate (cut the dimensions of) an infinite dimensional
Hilbert space into a finite dimension, and photon blockades \cite{miranowicz2013two,rabl2011photon}
which can be used as a tool for nonlinear optical-state truncation
(i.e., as a nonlinear quantum scissors) \cite{miranowicz2013two}.
However, to the best of our knowledge, no effort has yet been made to investigate
the higher order nonclassical properties of QCSs.
Motivated by the above facts, in addition to the conventional lower-order nonclassical properties, here we also aim to investigate higher order
nonclassical properties of two QCSs. The idea of the first type of QCS which is usually referred to as the nonlinear QCS was developed by \cite{buvzek1992coherent,miranowicz1994coherent} using one of the definitions of the infinite dimensional coherent state. Specifically, this type of QCSs were prepared by applying a truncated
displacement operator on the vacuum state 
 as follows
\begin{eqnarray}
|\alpha\rangle_{d}=\hat{D}_{d}(\alpha,\alpha^{*})|0\rangle & = & \exp(\alpha\hat{a}_{d}^{\dagger}-\alpha^{*}\hat{a}_{d})|0\rangle,\label{NCS}
\end{eqnarray}
where the truncated displacement operator $\hat{D}_{d}(\alpha,\alpha^{*})$
operates on vacuum to generate QCS, and the qudit annihilation operator
is $\hat{a}_{d}=\sum_{n=1}^{d-1}\sqrt{n}|n-1\rangle\langle n|$ and
the corresponding commutation relation is $[\hat{a}_{d},\,\hat{a}_{d}^{\dagger}]=d|d-1\rangle\langle d-1|$
which fundamentally differs from the standard creation and annihilation
operators. 
The Fock-state expansion of the QCS in the form of Eq. (\ref{eq:state1})
is given by \cite{miranowicz1994coherent}
\begin{equation}
|\alpha\rangle_{d}=\sum_{n=0}^{d-1}c_{n}^{(d)}(\alpha)|n\rangle,\label{NQCS}
\end{equation}
where the superposition coefficients are 
\begin{eqnarray}
c_{n}^{(d)}(\alpha) & =f_{n}^{(d)} & \sum_{k=0}^{d-1}\frac{{\rm He}_{n}(x_{k})}{[{\rm He}_{d-1}(x_{k})]^{2}}\exp(ix_{k}|\alpha|),\label{c_n}
\end{eqnarray}
with $f_{n}^{(d)}=\frac{(d-1)!}{d}(n!)^{-1/2}\exp[in\left(\phi_{0}-\tfrac{\pi}{2}\right)],$
and the modified Hermite polynomial ${\rm He}_{n}(x)$ is related
to the Hermite polynomial $H_{n}(x)$ as ${\rm He}_{n}(x)=2^{-n/2}H_{n}\left(x/\sqrt{2}\right)$;
$x_{k}\equiv x_{k}^{(d)}$ is the $k$ th root of ${\rm He}_{d}(x)$,
and $\phi_{0}={\rm arg}(\alpha)$. In the rest of the paper,
we have chosen $\phi_{0}=0$. The complex parameter $\alpha\left({\rm with}\,\phi_{0}=0\right)$
is perfectly periodic in nature for $d=2,\,3$ and almost periodic for
$d>3$. The periods of $\alpha$ for $d=2$ and $3$ are $T_{2}=\pi$
and $T_{3}=2\pi/\sqrt{3}$, respectively, whereas the periods for $d>3$
are $\sqrt{4d+2}$. Due the periodic nature of $\alpha$ the photon
number for the QCS $|\alpha\rangle$ is also periodic in nature with
maximum value $|\alpha|^{2}=d-1$ which corresponds to the photon number of the
highest energy Fock state. In Ref. \cite{miranowicz2014phase}
and references therein, possible ways of generating  this QCS
and a set of its nonclassical properties have been discussed. However, no attention
has yet been provided to the higher order nonclassical properties
of this state.

The second type of QCSs studied here can be generated by truncating the Fock space superposition of CS. This type of QCSs for a complex amplitude $\beta$
are defined as \cite{miranowicz2014phase,kuang1993dynamics,kuang1994coherent}
\begin{equation}
|\beta\rangle_{d}={\cal \mathscr{N}}\exp(\beta\hat{a}_{d}^{\dagger})|0\rangle={\cal \mathscr{N}}\sum_{n=0}^{d-1}\frac{\beta^{n}}{\sqrt{n!}}|n\rangle,\label{LCS-1}
\end{equation}
where $\mathcal{{\cal \mathscr{N}}}=1/(\sum_{n=0}^{d-1}{\cal }\frac{\beta^{2n}}{n!})^{1/2}$ is the normalization constant. This type of QCS is referred to as the linear QCS. The QCS  $|\beta\rangle_{d}$ can be written in the
form of Eq. (\ref{eq:state1}) with 
\begin{equation}
c_{n}^{d}\left(\beta\right)={\cal {\cal \mathscr{N}}}\frac{\beta^{n}}{\sqrt{n!}}.\label{C_n_Beta}
\end{equation}
This QCS is referred to as linear QCS \cite{miranowicz2014phase}
and was studied earlier in Refs.\cite{miranowicz2014phase,kuang1993dynamics,kuang1994coherent}.
In Ref. \cite{miranowicz2014phase}, it is explicitly shown that nonclassical
properties (e.g., Wigner function and nonclassical volume) of linear
QCS and nonlinear QCS are different. Here, we aim to extend the observation
further by comparing the nonclassical properties of these QCSs using
various witnesses and measures of nonclassicality with a specific
focus on the witnesses of higher order nonclassicality.
The remaining part of the paper is organized as follows. In Section
\ref{sec:COMPARISON-OF-NONCLASSICALITY-WITNESSES}, we compare nonclasscial
characters of linear and nonlinear QCSs using a set of witnesses of
nonclassicality which generally reflects the presence of higher order
nonclassicality (except Klyshko's criterion), but does not provide
any quantitative measure of nonclassiclaity. Specifically, in this
section, we perform comparison of nonclassicality in QCSs using the
criteria of higher order antibunching (HOA), higher order sub-Poissonian photon statistics (HOSPS), higher order squeezing (HOS) and Agarwal-Tara criterion and Klyshko's
criterion. In Section \ref{sec:Measure-of-nonclassicality}, we compare
the amount of nonclassicality present in linear and nonlinear QCSs
by using a set of quantitative measures of nonclassiclaity (e.g., concurrence
potential, negativity potential, and anticlassicality). Finally, the
paper is concluded in Section \ref{sec:CONCLUSION}.

\section{COMPARISON OF NONCLASSICALITY IN QCSs USING THE WITNESSES OF NONCLASSICALITY\label{sec:COMPARISON-OF-NONCLASSICALITY-WITNESSES} }
A quantum state is referred to as nonclassical if its Glauber Sudarshan
$P$-function cannot be written like a classical probability distribution.
In other words, negative values of $P$-function implies that the
state does not have classical analogue, and can be referred to as
a nonclassical state. As there does not exist any general procedure
for the measurement of $P$-function, several operational criteria
are designed to identify the signatures of nonclassicality. Most of
these criteria are only sufficient in the sense that if one of the
criteria is satisfied then the state is definitely nonclassical, but
if it's not satisfied then we cannot conclude anything about the nonclassicality
of the state. For example, in this section, we would discuss nonclassical
properties of linear and nonlinear QCSs using the criteria of HOA, HOS, HOSPS, Agarwal-Tara
and Klyshko's criteria. All these criteria are only sufficient. Further,
they are only witnesses of nonclassicality (only provides the signature
of nonclassicality) as none of them provide any quantitative measures
of the amount of nonclassicality present in a state. Of course, there
exist a handful of measures of nonclassicality, but each of them has
some limitations (see \cite{miranowicz2015statistical} for a discussion).
We will get back to the issues of nonclassicality measures in the
next section, where we will compare linear and nonlinear QCSs using
some of those measures. This section is focused on witnesses of nonclassicality
and in what follows we would compare nonclassical features of linear
and nonlinear QCSs using the criteria of HOA, HOS, HOSPS and Agarwal-Tara
and Klyshko's criteria. 
All these criteria  are based on 
moments of photon number and/or quadrature and are thus experimentally
measurable with the general measurement scheme proposed in Ref. \cite{shchukin2005nonclassical}.
It may be further noted that an infinite set (or hierarchy) of
such moment based criteria is equivalent to the $P$-function, i.e.,
is both necessary and sufficient in nature \cite{richter2002nonclassicality}.
Recently, Miranowicz et al., have summarized all the existing nonclassicality
criteria for both single- and multi-mode bosonic fields and proposed
a unified approach to generate new operational inequalities to characterize
nonclassicality in the radiation \cite{miranowicz2010testing}. In what follows, we would use a small set of these nonclassicality witnesses
to perform the proposed comparison. To begin with, we would like to
discuss the possibility of observing HOA in linear and nonlinear QCSs.
\subsection{Higher order antibunching}
\begin{figure}
\centering{}
\subfigure[]{\includegraphics[scale=0.6]{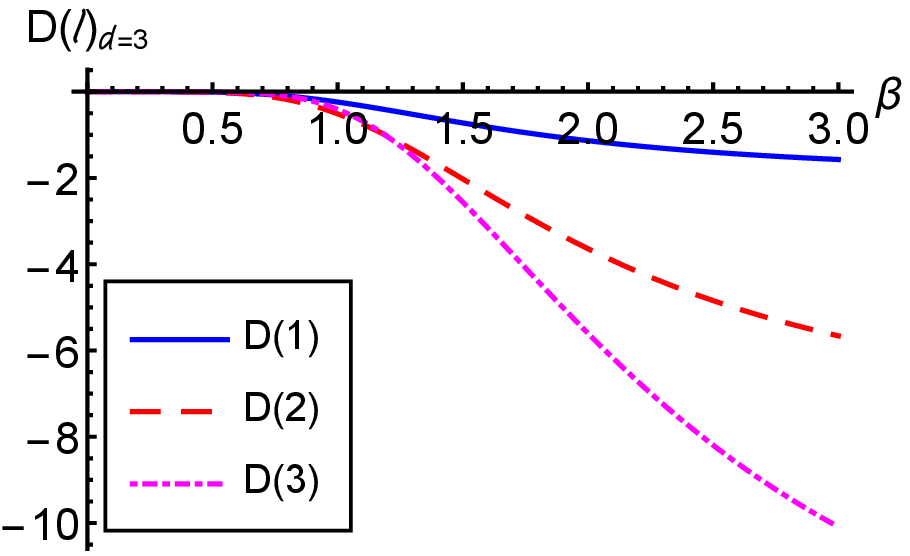}}\quad\quad \subfigure[]{\includegraphics[scale=0.6]{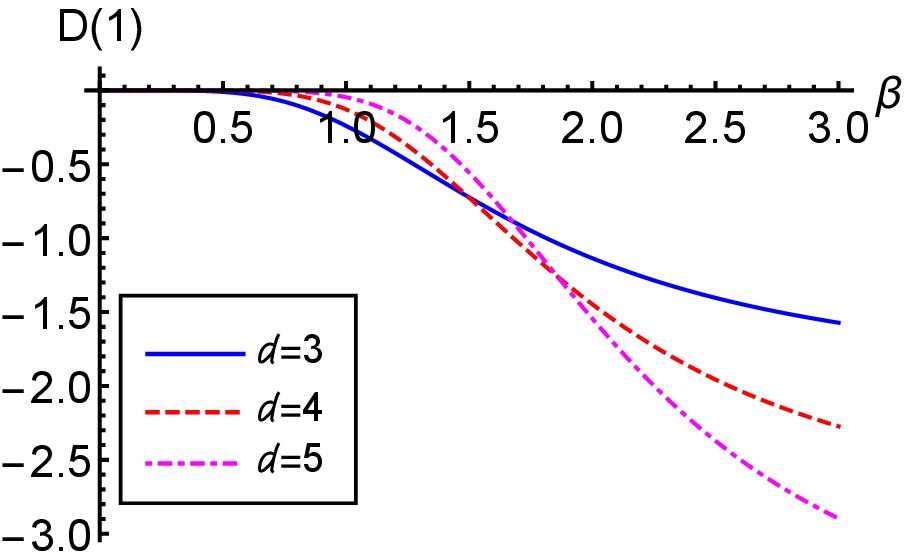}}\\
\subfigure[]{\includegraphics[scale=0.6]{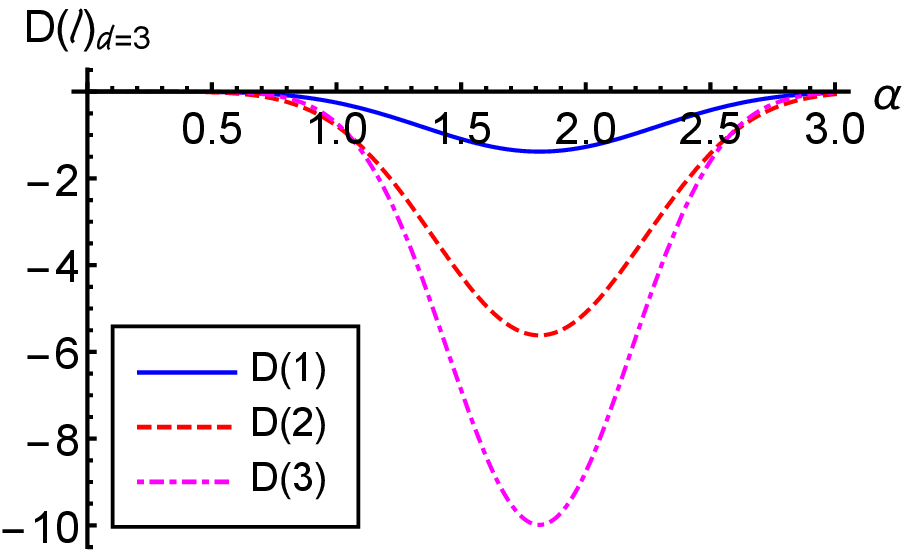}}\quad\quad  \subfigure[]{\includegraphics[scale=0.6]{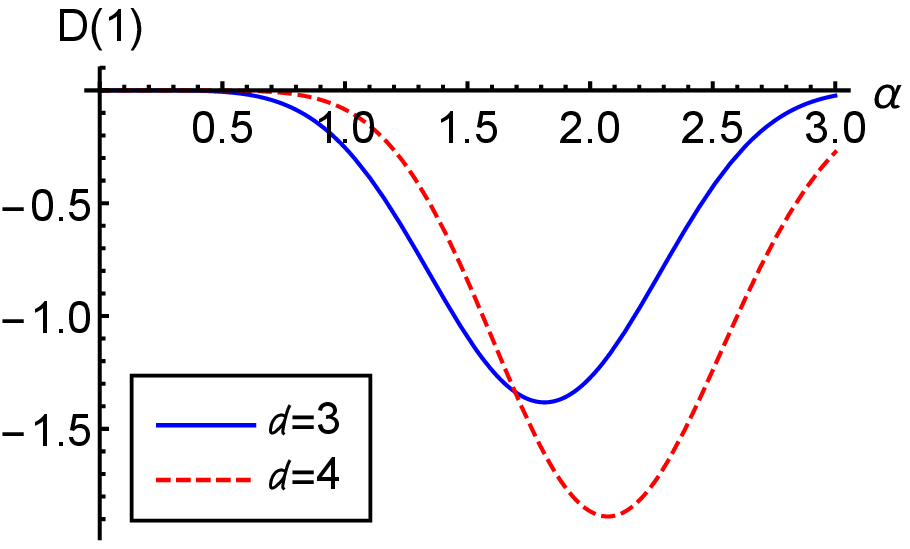}}\\
\caption{\label{fig:HOA}(Color online) The variation of higher order antibunching
is shown with state parameter $\alpha$ or $\beta$. In (a), higher order
antibunching in state $\left|\beta\right\rangle _{d=3}$ is shown with
$l=1$ (smooth blue line), $l=2$ (dashed red line) and $l=3$ (dotted
dashed magenta line). (b) shows the effect of change of dimension
$d$ of finite dimensional state $\left|\beta\right\rangle _{d}$
for $d=3,\,4$ and $5$ with smooth (blue), dashed (red) and dotted
dashed (magenta) lines, respectively. (c) and (d) show the similar
variation for state $\left|\alpha\right\rangle _{d}$. (c) shows higher
order antibunching for dimension $d=3$ with $l=1$ (smooth blue line),
$l=2$ (dashed red line) and $l=3$ (dotted dashed magenta line).
(d) shows the usual antibunching in $\left|\alpha\right\rangle _{d}$
for dimensions $d=3$ and $4$ with smooth (blue) and dashed (red)
lines, respectively.}
\end{figure}

Different well-known criteria of higher order nonclassicality can
be expressed in compact forms for the finite dimensional states given
by Eq. (\ref{eq:state1}). In this section, we would focus on HOA.
The concept of HOA was introduced by Lee in 1990 \cite{lee1990many}.
In fact, he provided a criterion for HOA using the
theory of majorization. Subsequently, in 2002, Lee's criterion was
modified by Ba An \cite{an2002multimode}, who introduced
a simplified criterion for HOA and used that to show the existence
of HOA in the trio coherent state. Later, in 2006, Pathak and Garcia \cite{pathak2006control}
provided a clear physical meaning to HOA and further simplified the
criterion of HOA. This criterion is now known as Pathak-Garcia criterion,
and in what follows, we use Pathak-Garcia criterion \cite{pathak2006control}
of HOA as a witness of nonclassicality. According to this criterion,
$l$th order antibunching is observed if following inequality is satisfied
by a quantum state
\begin{equation}
D(l)=\left\langle N^{(l+1)}\right\rangle -\left\langle N\right\rangle ^{l+1}=\langle a^{\dagger(l+1)}a^{(l+1)}\rangle-\langle a^{\dagger}a\rangle^{l+1}<0.\label{eq:ho21}
\end{equation}
\textcolor{black}{For }$l=1,$ \textcolor{black}{one can obtain a usual
antibunching criterion. Before proceeding further, we would like to
note that until the recent past, HOA was supposed to be a very rare phenomenon,
but in 2006, some of the present authors have established that HOA
is not so rare, and} subsequently HOA has been reported in various
infinite dimensional quantum states (e.g., nonlinear squeezed state
\cite{verma2010generalized}, photon-added coherent state \cite{verma2008higher},
photon added and subtracted squeezed coherent state \cite{thapliyal2017comparison})
and a set of finite dimensional states (e.g., binomial state \cite{verma2010generalized},
generalized binomial state \cite{verma2008higher}, reciprocal binomial
state \cite{verma2008higher}, negative binomial state \cite{verma2008higher},
hypergeometric state \cite{verma2008higher}). It has also been reported
in a set of physical systems (e.g., in asymmetric
nonlinear optical coupler \cite{thapliyal2014higher,thapliyal2014nonclassical},
BEC systems \cite{giri2014single,giri2017nonclassicality}, coupled anharmonic oscillators \cite{alam2015approximate,alam2016nonclassical,alam2016quantum} an optomechanical
and optomechanics-like systems \cite{alam2017lower}, Raman \cite{sen2007quantum}
and hyper-Raman \cite{sen2007squeezing} processes).
However, to the best of our knowledge, no effort has yet been made
to investigated the possibilities of observing HOA in QCSs. Keeping
this in mind, in this section, we investigate the possibilities of
observing HOA in linear and nonlinear QCSs.
For the finite dimensional states of the form (\ref{eq:state1}),
we can have compact expressions for higher order antibunching using
Eqs. (\ref{eq:ho21}) as follows \cite{verma2010generalized} 
\begin{equation}
D(l)=\left\langle N^{(l+1)}\right\rangle -\left\langle N\right\rangle ^{l+1}=\sum_{j=0}^{d-1}\frac{j!}{(j-l-1)!}|c_{j}|^{2}-\left(\sum_{j=0}^{d-1}|c_{j}|^{2}j\right)^{l+1}<0.\label{eq:HOA-simple}
\end{equation}
Now, we can obtain $D(l)$ for nonlinear and linear QCSs with the help of 
Eqs. (\ref{c_n}), (\ref{C_n_Beta}) and  (\ref{eq:HOA-simple}), respectively and the corresponding expressions
of $D(l)$s are plotted in Fig. \ref{fig:HOA} which clearly shows
the variation of higher order antibunching in QCS with parameter $\alpha$($\beta$) for nonlinear (linear) QCSs. All the plots in Fig. \ref{fig:HOA}
show that the states are antibunched (of different orders) for the
particular values of the parameters used here. Specifically, Figs.
\ref{fig:HOA}(a) and \ref{fig:HOA}(c) show that the depth of nonclassicality
increases with the order of antibunching in linear and nonlinear QCSs,
respectively. Similarly, Figs. \ref{fig:HOA}(b) and  \ref{fig:HOA}(d) show the
effect of dimensions on antibunching in linear and nonlinear QCSs,
respectively.
\subsection{Higher order squeezing}
\begin{figure}
\begin{centering}
\subfigure[]{\includegraphics[scale=0.5]{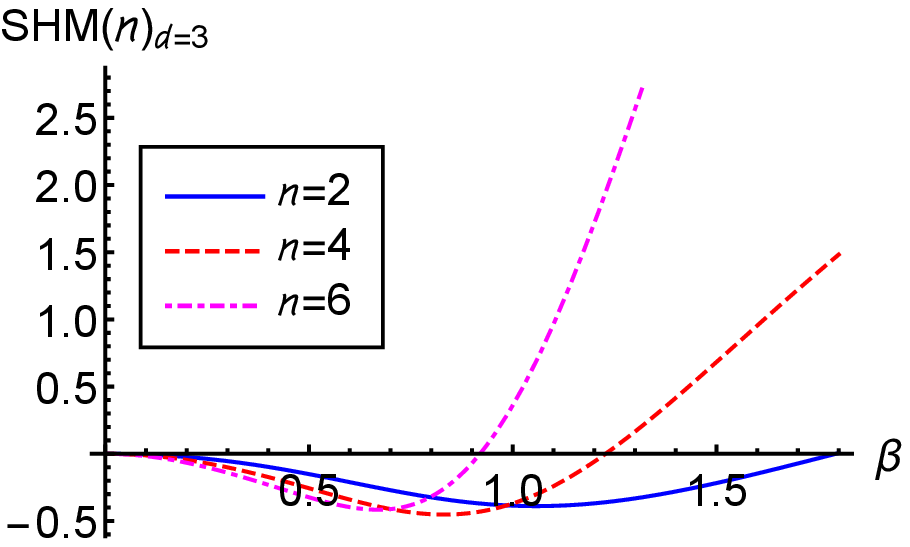}}\quad\quad \subfigure[]{\includegraphics[scale=0.5]{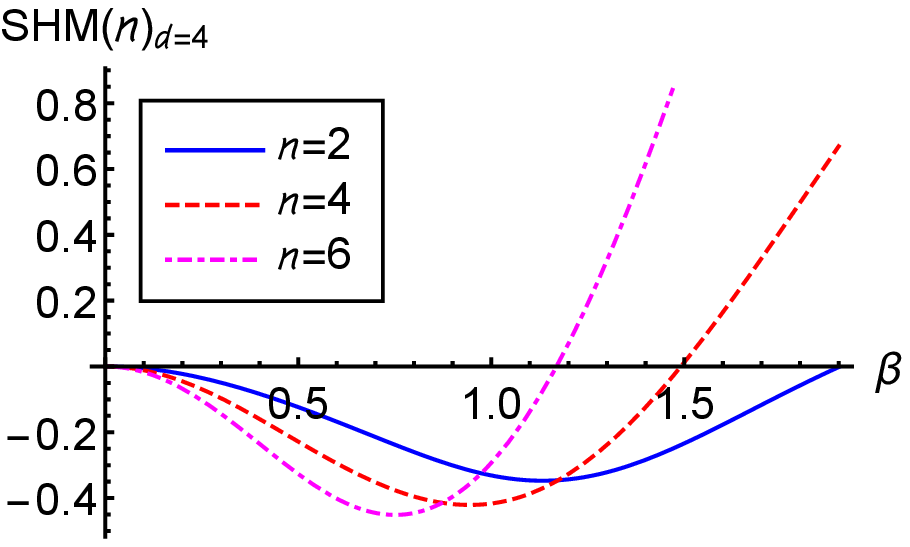}} \quad\quad \subfigure[]{\includegraphics[scale=0.5]{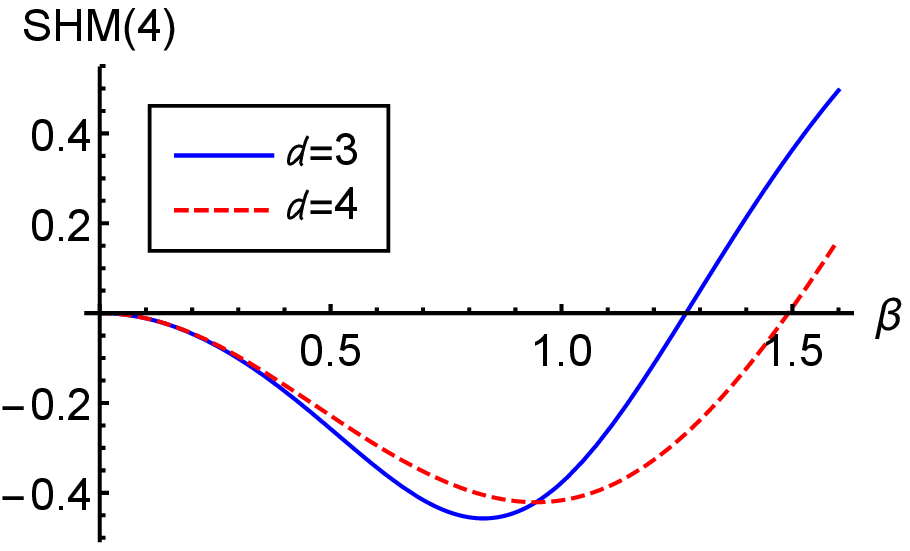}}\\
\subfigure[]{\includegraphics[scale=0.5]{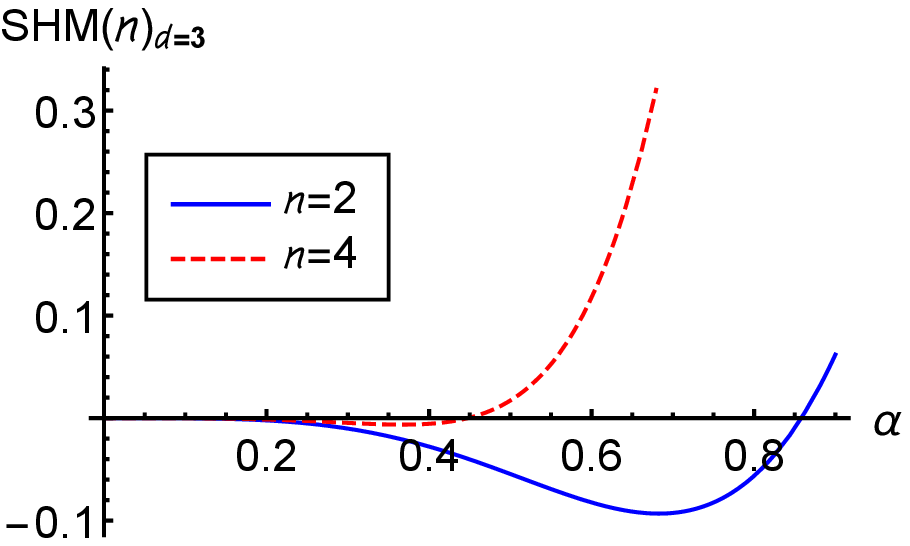}}\quad\quad \subfigure[]{\includegraphics[scale=0.5]{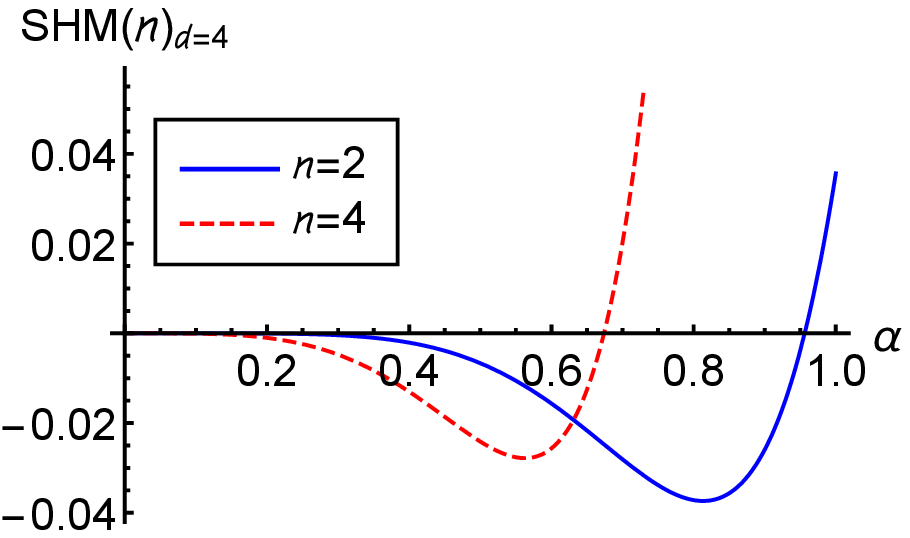}}\quad\quad \subfigure[]{\includegraphics[scale=0.5]{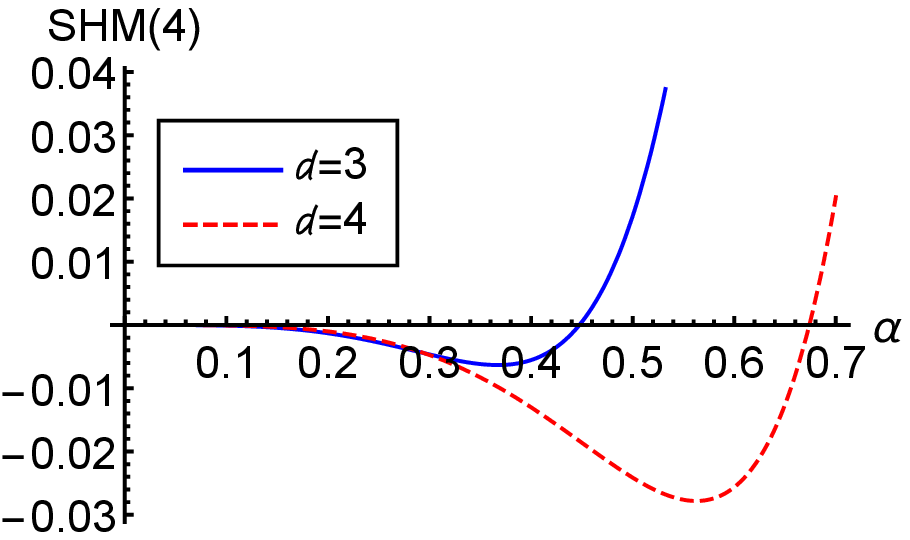}}\\
\end{centering}
\caption{\label{fig:SHM}(Color online) The variation of Hong-Mandel higher
order squeezing parameter is shown with the parameter $\alpha$ or
$\beta$. (a) and (b) show higher order squeezing in state $\left|\beta\right\rangle _{d}$
for $d=3$ and $4$, respectively, with $n=2$ (smooth blue line),
$n=4$ (dashed red line) and $n=6$ (dotted dashed magenta line).
Similarly, (d) and (e) show higher order squeezing in state $\left|\alpha\right\rangle _{d}$
for $d=3$ and $4$, respectively, with $n=2$ (smooth blue line),
$n=4$ (dashed red line). (c) and (f) show the Hong-Mandel squeezing
for $n=4$ with $d=3$ (smooth blue line) and $d=4$ (dashed red line)
for states $\left|\beta\right\rangle _{d}$ and $\left|\alpha\right\rangle _{d}$,
respectively.}
\end{figure}

\textcolor{black}{Further, higher order squeezing can be studied using two independent
crite}ria: Hillery's criterion
for amplitude powered squeezing \cite{hillery1987amplitude}
and Hong-Mandel's criterion \cite{hong1985higher}. Here, we investigate
higher order squeezing using Hong-Mandel's criterion which can be
described by the following inequality
\begin{equation}
\langle(\Delta X)^{n}\rangle=\sum_{r=0}^{n}\sum_{i=0}^{\frac{r}{2}}\sum_{k=0}^{r-2i}(-1)^{r}\frac{1}{2^{\frac{n}{2}}}(2i-1)!!\,^{r-2i}C_{k}\,^{n}C_{r}\,^{r}C_{2i}\langle a^{\dagger}+a\rangle^{n-r}\langle a^{\dagger k}a^{r-2i-k}\rangle<\left(\frac{1}{2}\right)_{\frac{n}{2}}=\frac{1}{2^{\frac{n}{2}}}(n-1)!!,\label{eq:cond2.1}
\end{equation}
where $n$ is an even number, quadrature variable is defined as $X=\frac{1}{\sqrt{2}}\left(a+a^{\dagger}\right)$
and $(x)_{r}$ is conventional Pochhammer symbol.
Now, for the finite dimensional states of the form (\ref{eq:state1}),
we can have a compact expressions for HOS using Eqs. (\ref{eq:ho21})
and (\ref{eq:cond2.1}) in the form \cite{verma2010generalized} 
\begin{equation}
\begin{array}{lcl}
\langle(\Delta X)^{n}\rangle & = & \sum\limits _{r=0}^{n}\sum_{i=0}^{\frac{r}{2}}\sum_{k=0}^{r-2i}(-1)^{r}\frac{1}{2^{\frac{n}{2}}}(2i-1)!!\,^{r-2i}C_{k}\,^{n}C_{r}\,^{r}C_{2i}\\
 & \times & \left(\sum\limits _{m=0}^{N-1}\sqrt{(m+1}\left(c_{m}c_{m+1}^{*}+c_{m}^{*}c_{m+1}\right)\right)^{n-r}\\
 & \times & \sum\limits _{j=0}^{N-Max[k,\,r-2i-k]}c_{j+k}^{*}c_{j+r-2i-k}\frac{1}{j!}\left((j+r-2i-k)!(j+k)!\right)^{\frac{1}{2}}<\left(\frac{1}{2}\right)_{\frac{n}{2}}.
\end{array}\label{eq:gen4}
\end{equation}
Using Eq. (\ref{eq:gen4}) and the expressions of $c_{n}^{d}$ from Eqs.
(\ref{eq:HOA-simple}) and (\ref{C_n_Beta}), we investigate the existence of HOS in nonlinear and linear coherent
states. The same is performed through the corresponding plots shown
in Fig. \ref{fig:SHM}, which illustrates the existence of Hong-Mandel
type HOS in both linear and nonlinear QCSs. Figs. \ref{fig:SHM}(a)
and \ref{fig:SHM}(d) show the higher order squeezing in linear and nonlinear QCSs,
respectively, for $d=3$. Similarly, Figs. \ref{fig:SHM}(b) and \ref{fig:SHM}(e)
show higher order squeezing for $d=4$. While Figs. \ref{fig:SHM}(c) and \ref{fig:SHM}(f) compare the higher order squeezing with $n=4$ for different
dimensions ($d=3$ and $4$) for both linear and nonlinear QCSs, respectively.
Comparing Fig. \ref{fig:SHM}(a) with Fig. \ref{fig:SHM}(c), and
Fig. \ref{fig:SHM}(b) with Fig. \ref{fig:SHM}(d) we can easily
conclude that the higher order nonclassical properties of linear and
nonlinear coherent states are different.
\subsection{Higher Order sub-Poissonian Photon Statistics }
Higher order sub-Poissonian photon statistics \cite{lee1990higher,agarwal1992nonclassical}
is an important aspect of the study of the existence of higher order nonclassicality
and quantum statistical properties of the radiation field. In the
recent past, HOSPS has been reported in a set of infinite dimensional
states (e.g., nonlinear squeezed state \cite{verma2010generalized},
photon-added coherent state \cite{verma2008higher}, photon added
and subtracted squeezed coherent states \cite{thapliyal2017comparison})
and for a set of finite dimensional states (e.g., binomial state \cite{verma2010generalized}).
The generalized criterion for observing the $\left(l-1\right)$th
order sub-Poissonian photon statistics is given by
\cite{verma2010generalized}
\begin{equation}
\begin{array}{lcccc}
D_{h}(l-1) & = & \sum\limits _{r=0}^{l}\sum\limits _{k=0}^{r}S_{2}(r,\,k)\,^{l}C_{r}\,\left(-1\right)^{r}D(k-1)\langle N\rangle^{l-r} & < & 0,\end{array}\label{eq:hosps22}
\end{equation}
where $S_{2}(r,\,k)$ is the Stirling numbers of second kind. \textcolor{black}{In
order to obtain the analytical expression for HOSPS for the QCSs,
we use Eqs. (\ref{eq:state1}), (\ref{eq:HOA-simple}) and (\ref{eq:hosps22}) which after simplification yields }
\begin{equation}
\begin{array}{lcccc}
D_{h}(l-1) & = & \sum\limits _{r=0}^{l}\sum\limits _{k=0}^{r}S_{2}(r,\,k)\,^{l}C_{r}\,\left(-1\right)^{r}\left\{ \sum\limits _{j=0}^{d-1}\frac{j!}{(j-k-2)!}|c_{j}|^{2}-\left(\sum\limits _{j=0}^{d-1}|c_{j}|^{2}j\right)^{k}\right\} \left(\sum\limits _{j=0}^{d-1}|c_{j}|^{2}j\right)^{l-r} & < & 0.\end{array}\label{eq:hosps_final}
\end{equation}
For the nonlinear and linear QCSs, the analytic expressions for the coefficients $c_{j}$ are used 
from Eqs. (\ref{c_n})
and (\ref{C_n_Beta}), respectively. After performing few step algebra, we find out analytic expression for  $D_{h}(l-1)$. We  plot  $D_{h}(l-1)$ with respect
to $\alpha$ and $\beta$ for the different values of $l$, the result is depicted in Figs.
\ref{HOSPS_A3}(a) and \ref{HOSPS_A3}(c). The results shown in Figs. \ref{HOSPS_A3}(a) and \ref{HOSPS_A3}(c) illustrate the existence of HOSPS through
the presence of negative regions. Specifically, of the figure ensures
the existence of higher order sub-Poissonian photon statistics for
$l>1$ and sub-Poissonian photon statistics for $l=1$. Further, we
can see that for a linear coherent state the nonclassicality witness
for HOSPS is found to vary monotonically with $\beta$ and finally
to approach a saturation. However, in case of nonlinear QCS, the nonclassicality
witness is found to show a kind of oscillation with frequency same as $\alpha$.
\subsection{Agarwal-Tara Criterion\label{subsec:Agarwal-Tara-Criterion}}
\begin{figure}
\begin{centering}
\subfigure[]{\includegraphics[scale=0.6]{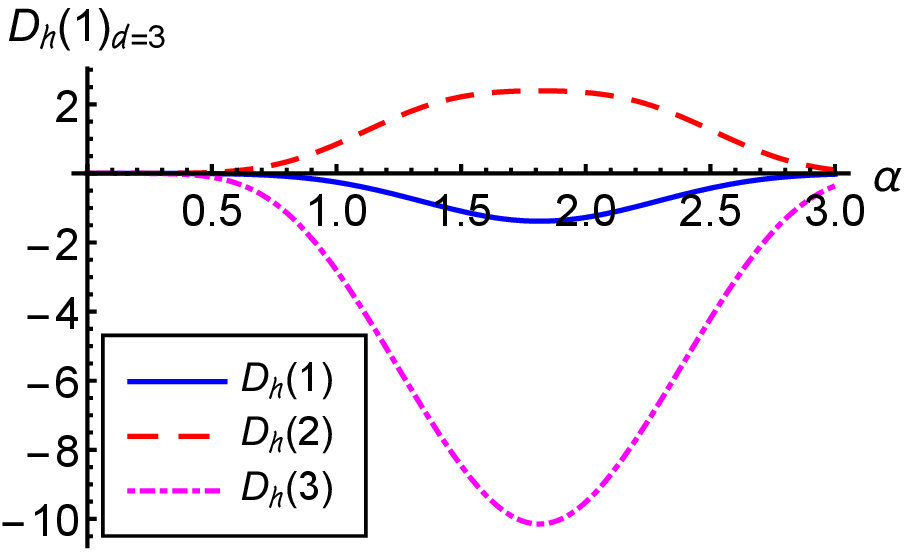}}\quad\quad \subfigure[]{\includegraphics[scale=0.6]{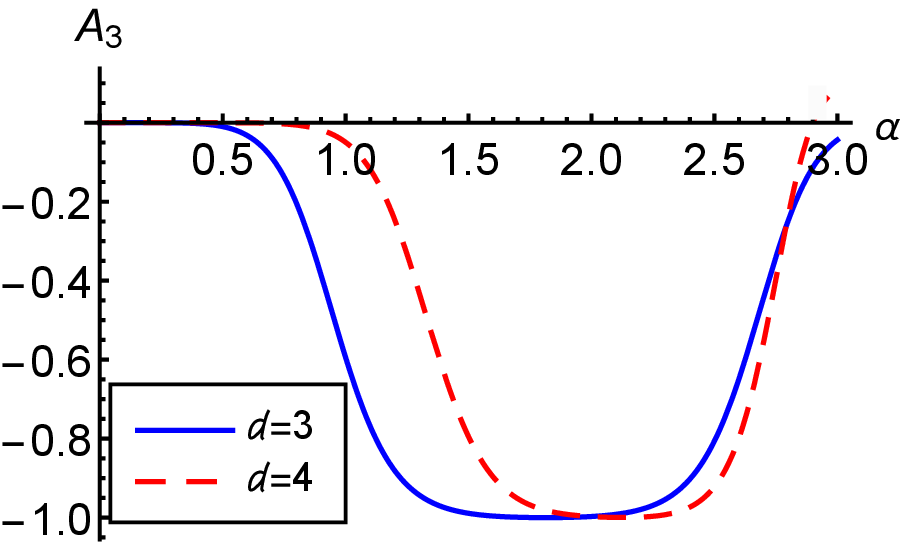}}\\
\subfigure[]{\includegraphics[scale=0.6]{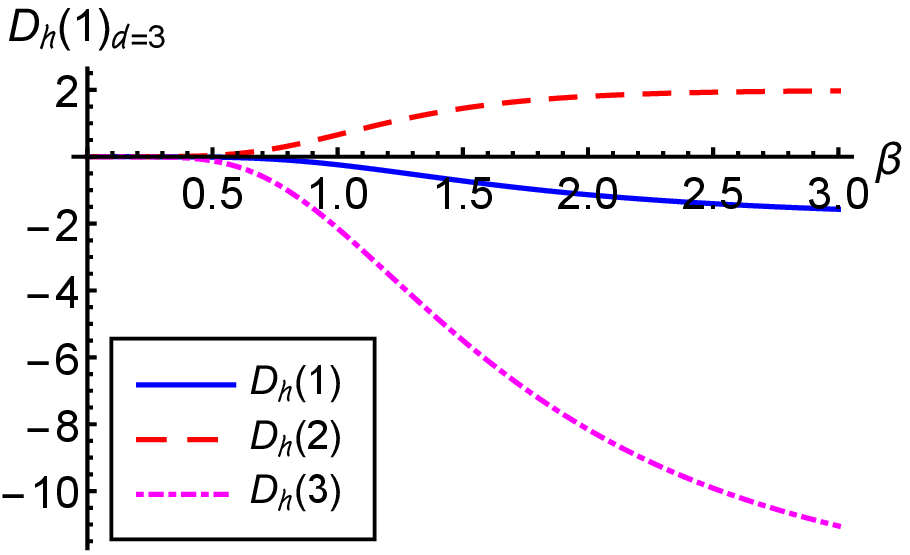}}\quad\quad \subfigure[]{\includegraphics[scale=0.6]{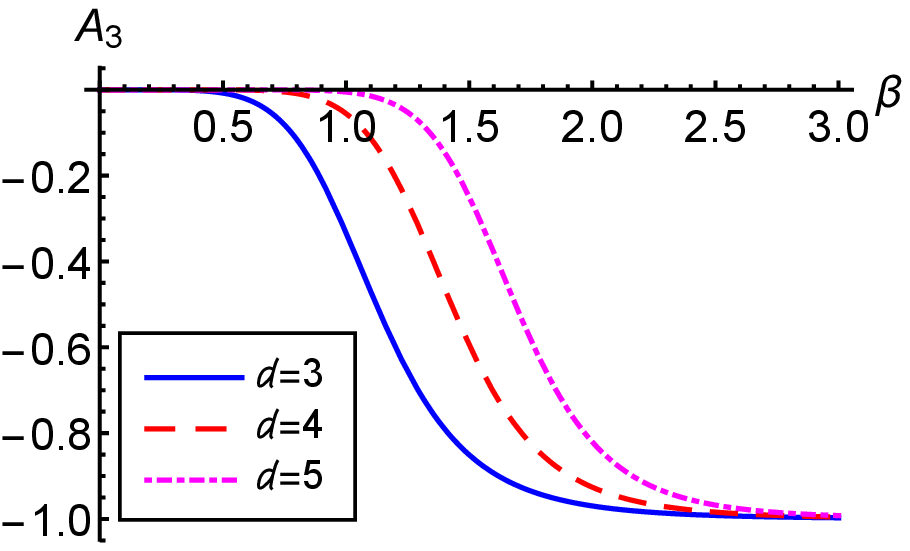}}\\
\end{centering}
\caption{\label{HOSPS_A3}The variation of $D_{h}(l)$ parameter illustrating HOSPS and $A_{3}$ parameter 
for nonclassicality criterion are shown with the parameter $\alpha$
or $\beta$. (a) and (c) show HOSPS in state $\left|\alpha\right\rangle _{d}$
and $\left|\beta\right\rangle _{d}$ for $d=3$ with $l=2$ (smooth
blue line), $l=3$ (dashed red line) and $l=4$ (dotted dashed magenta
line). (b) and (d) show $A_{3}$ criterion in state $\left|\alpha\right\rangle _{d}$
and $\left|\beta\right\rangle _{d}$ for $d=3$ (smooth blue line),
$d=4$ (dashed red line) and $d=5$ (dotted dashed magenta line). }
\end{figure}

In Ref. \cite{agarwal1992nonclassical}, Agarwal and Tara introduced
a moment based criterion to investigate the  witness of the nonclassical characteristics of a
given quantum state. They introduced a parameter $A_{3}$ which consists
of the moments $\left(\mu_{n}=\langle\left(a^{\dagger}a\right)^{n}\rangle\right)$
of the number distribution and the normal ordered moments $\left(m_{n}=\langle a^{\dagger n}a^{n}\rangle\right)$.
The analytic expression of $A_{3}$ in terms of these moments and
condition for nonclassicality is given by 
\begin{equation}
A_{3}=\frac{\det\,m^{(3)}}{\det\,\mu^{(3)}-\det\,m^{(3)}}<0,\label{eq:A3}
\end{equation}
where,
\[
\begin{array}{ccc}
m^{(3)}=\left[\begin{array}{ccc}
1 & m_{1} & m_{2}\\
m_{1} & m_{2} & m_{3}\\
m_{2} & m_{3} & m_{4}
\end{array}\right] & {\rm and} & \mu^{(3)}=\left[\begin{array}{ccc}
1 & \mu_{1} & \mu_{2}\\
\mu_{1} & \mu_{2} & \mu_{3}\\
\mu_{2} & \mu_{3} & \mu_{4}
\end{array}\right]\end{array}
\]
The  normal order and number distribution moments for  finite dimensional states are calculated using Eq. (\ref{eq:state1}). The analytic expressions for the corresponding moments are given by 
\begin{equation}
\begin{array}{lcl}
\langle a^{\dagger n}a^{n}\rangle & = & \sum\limits _{j=0}^{d-1}\frac{j!}{(j-n)!}|c_{j}|^{2},\\
\langle\left(a^{\dagger}a\right)^{n}\rangle & = & \sum\limits _{j=0}^{d-1}\sum\limits _{k=0}^{n}S_{2}(n,k)\frac{j!}{(j-k)!}|c_{j}|^{2}.
\end{array}\label{A3_moments}
\end{equation}
The parameter $A_{3}$ is zero for coherent state (classical state)
and $-1$ for a Fock state (most nonclassical state), respectively.
For a nonclassical state, $A_{3}$ is negative and bounded by the
value -1 when the state became maximally nonclassical state. In order
to investigate $A_{3}$ for both types of QCSs, we use the expression of $c_{n}^{(d)}(\alpha)$
and $c_{n}^{(d)}\left(\beta\right)$, and
Eq. (\ref{A3_moments}). After calculating the analytic expression of  inequality for the  $A_{3}$, we plot the corresponding results for the
QCSs. In Figs. \ref{HOSPS_A3}(b) and \ref{HOSPS_A3}(d), we have
shown the result where we have observed that for both linear and nonlinear
QCSs $A_{3}$ varies between zero to $-1$ with respect to $\alpha$
and $\beta$, and thus depict the presence of noclassicality. The periodic nature of $A_{3}$ with respect to $\alpha$ and $\beta$ is observed to be similar with that for the witness of HOSPS. The zero value of $A_{3}$
for $\alpha$ ($\beta$) in Fig. \ref{HOSPS_A3}(b) (\ref{HOSPS_A3}(d))
is consistent with the fact that vacuum state is a classical state
having non-negative $P$-function. 
\subsection{Klyshko's criterion}
\begin{figure}
\centering
\subfigure[]{\includegraphics[scale=0.4]{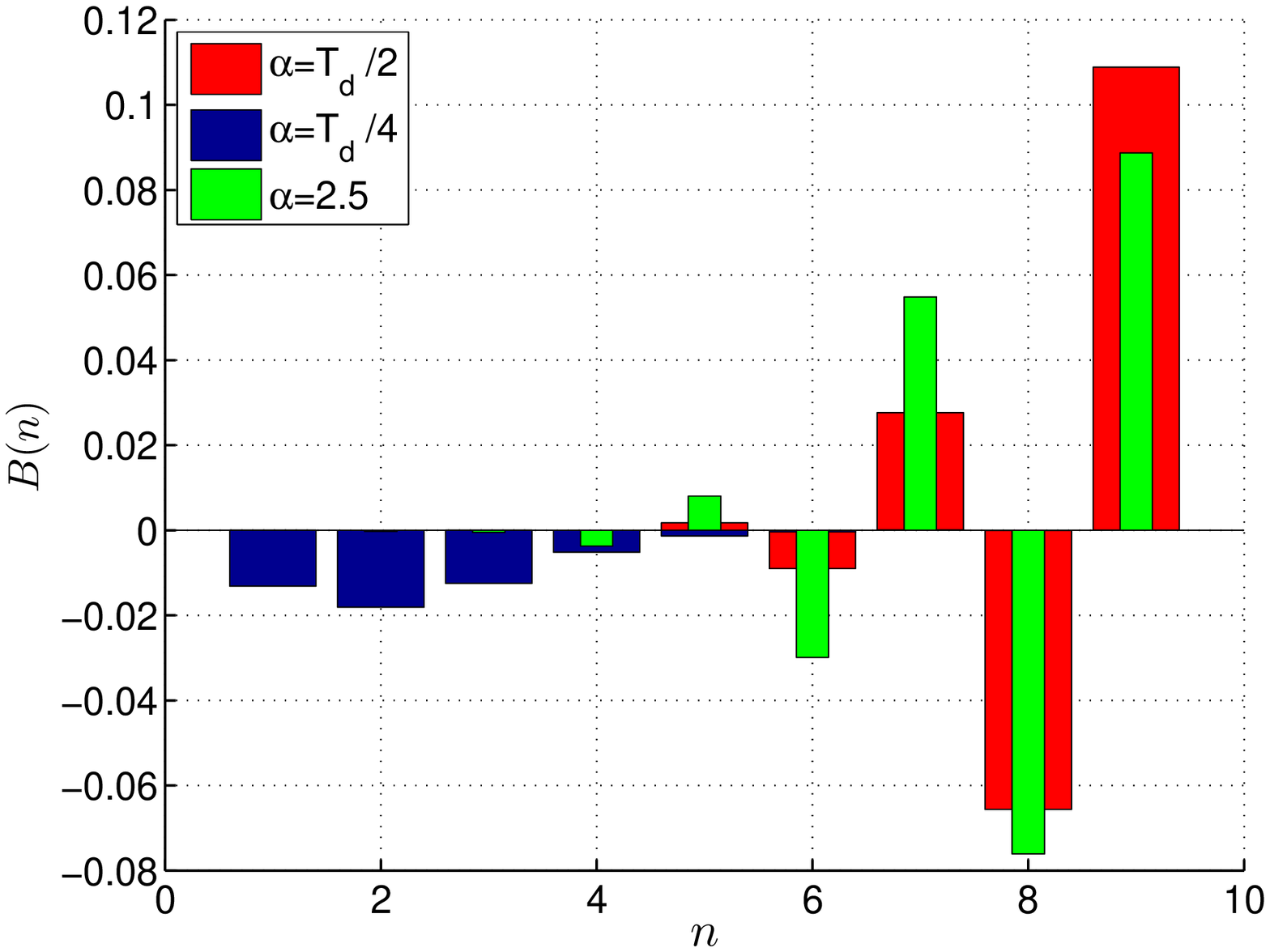}}\quad\quad \subfigure[]{\includegraphics[scale=0.4]{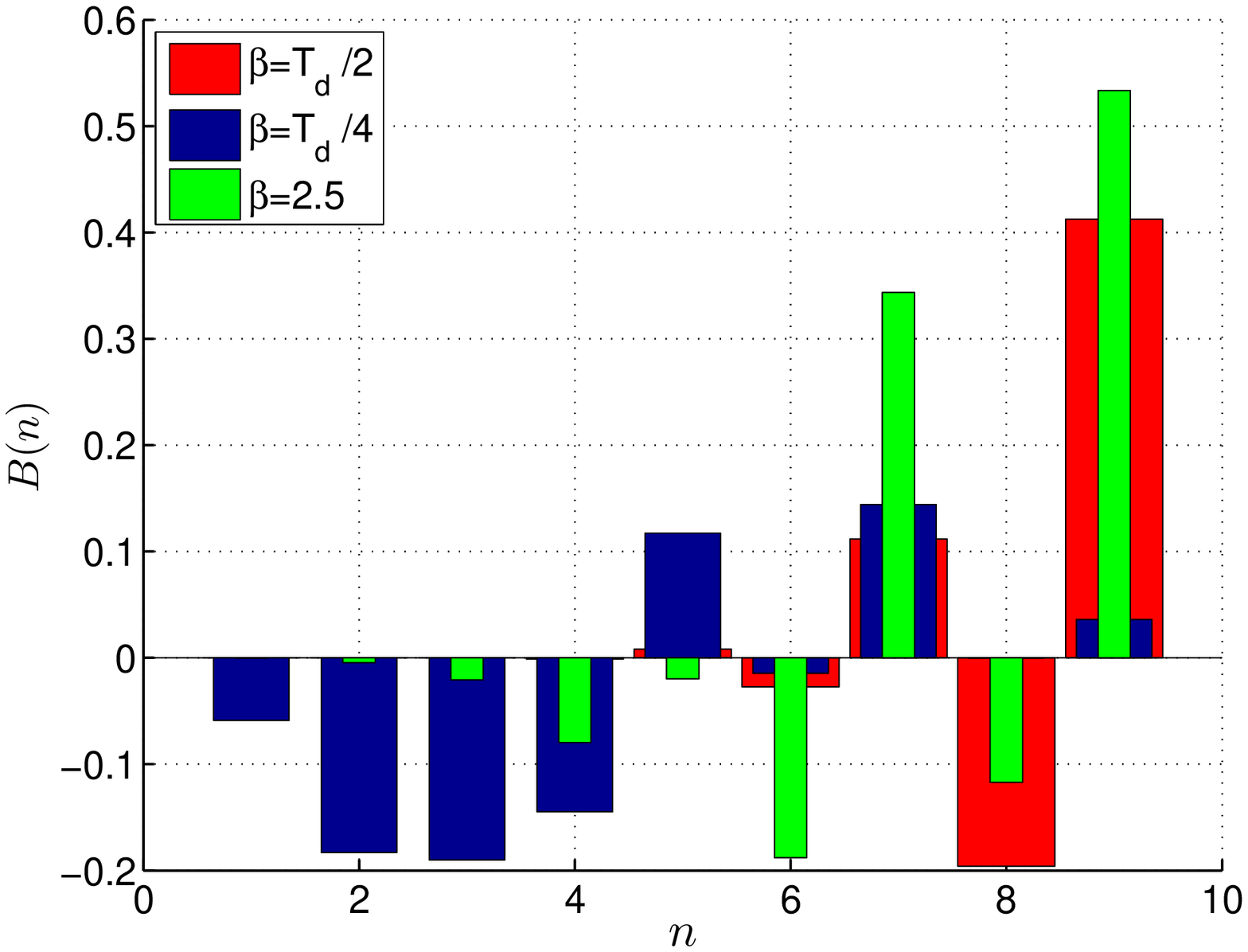}}\\
\caption{\label{fig:Illustration-of-Klyshko's} Illustration of Klyshko's criterion (a) variation of $B(n)$ with respect to  $n$ for QCS $|\alpha\rangle_{d}$ where thick (red) bar, medium (blue) bar and thin (green) bar  correspond  to $0.06\times B(n)$, $B(n)$ and $0.5\times B(n)$ with $\alpha=T_{d}/2, T_{d}/4$ and 2.5, respectively, and (b) variation of   $B(n)$  with respect to $n$ for QCS $|\beta\rangle_{d}$ where thick (red) bar, medium (blue) bar and thin (green) bar correspond
to $\beta=T_{d}/2, T_{d}/4$ and 2.5, respectively. }
\end{figure}

Klyshko \cite{klyshko1996nonclassical} introduced a nonclassicality
witness involving probability, $p_{n}=\langle n|\rho|n\rangle$ of
obtaining Fock state $|n\rangle$, as follows
\begin{equation}
B\left(n\right)\equiv\left(n+2\right)p_{n}\,p_{n+2}-\left(n+1\right)\left[p_{n+1}\right]^{2}<0.\label{eq:Klyshko}
\end{equation}
This inequality has an advantage over other existing criteria, as in this criteria we need only the photon number distribution $p(n)$ for the three successive values of $n$, whereas other criteria a complete description is required. Further,
this criterion has been recently employed to reveal nonclassicality
present in various quantum states \cite{thapliyal2017comparison}.
Here, we observe $B\left(n\right)$ to be negative for both linear
and nonlinear QCS which indicate the existence of a nonclassical photon
statistics. Such signatures of nonclassical photon statistics have
already been found through the investigation on HOA and HOSPS. However, the 
satisfaction of any of them (HOA and HOSPS) does not warrant the
satisfaction of Klyshko's criterion. Further, for a nonlinear QCS,
detectability of the nonclassical character via Klyshko's criterion
is found to depend on the dimension of the state and choice of $\alpha.$ We calculate the inequality $B(n)$ for the both types of the QCSs with the help of       Eqs. (\ref{NQCS}), (\ref{c_n}), (\ref{C_n_Beta}) and (\ref{eq:Klyshko}), respectively. 
In Figs. \ref{fig:Illustration-of-Klyshko's}(a) and \ref{fig:Illustration-of-Klyshko's}(b),
we clearly visualize the Klyshko's criteria for QCSs $|\alpha\rangle$
and $|\beta\rangle$ for different values of $\alpha$ and $\beta$
where the negative values of the $B(n)$, indicates the existence
of the nonclassicality. It is noticeable that $B(n)$ is negative
for particular Fock dimension for the given values of $\alpha$ and
$\beta$. 

\section{Measures of nonclassicality\label{sec:Measure-of-nonclassicality} }
So far we have discussed the possibilities of observing nonclassicality
in QCS using a set of witnesses of nonclassicality. In the process,
we have tried to compare the nonclassical nature of linear and nonlinear
coherent states, but such a comparison was only qualitative in nature.
To perform a quantitative comparison, we would require one or more
quantitative measures of nonclassiclaity. In fact, many quantitative
measures of nonclassicality (e.g., nonclassical distance, nonclassical
depth, negative volume of Wigner function, negativity potential, concurrence potential)
are in existence, but each of them have their own limitations (see
\cite{miranowicz2015statistical} for a review). For example, Hillery
introduced a measure of nonclassicality, nonclassical distance \cite{hillery1987nonclassical},
as trace norm distance between the state under consideration and closest
classical state. Computation of such a measure has drawback in requirement
of optimization over infinite number of parameters. Subsequently,
Lee \cite{lee1991measure} proposed another universal measure of nonclassicality known as nonclassical
depth which is equal to the amount of Gaussian noise required to transform
corresponding $P$-function into a classical probability distribution
function. It's established that this measure is not useful for non-Gaussian pure states \cite{lutkenhaus1995nonclassical}. The volume of the negative part of the Wigner function is also used as a quantitative
measure of nonclassicality, known as nonclassical volume \cite{kenfack2004negativity}.
However, this measure reflects the drawbacks of the Wigner function,
which fails to detect nonclassicality present in the Gaussian states (especially,
squeezed states). 

With the advancement of the quantum information and computation, various
measures of entanglement (such as negativity, concurrence) have been
introduced. In 2005, Asobth \cite{asboth2005computable} introduced
an excellent idea to use the measures entanglement for quantifying nonclassicality present
in a quantum state. This whole idea relies on the conjecture that linear
optics conserves nonclassicality \cite{vogel2014unified,ge2015conservation},
therefore, it maps the amount of nonclassicality in the input ports
of the beam splitter (BS) to the same amount of entanglement at its output
ports. Thus, if we send the vacuum state through one of the input ports and a single mode nonclassical state through the other port of the BS, the amount
of bipartite entanglement present in the output ports would quantify the amount
of single mode nonclassicality in the quantum state under consideration.
Note that if a classical state is inserted in the same manner, then
the output will always be separable. This observation, guided Asboth
to introduce a measure of the nonclassicality of the input single
mode state, which was referred to as entanglement potential. In the
original paper \cite{asboth2005computable}, logarithmic negativity
and entropic entanglement potential are used as entanglement measures,
but there exists a number of entanglement measures. In principle,
one of these entanglement measures can be used to quantify nonclassicality.
Here, we use entanglement potential as measure of single mode nonclassicality.
For example, if logarithmic negativity (concurrence) is used as the
measure of entanglement in the output ports, then we would refer to
it as negativity (concurrence) potential. In what follows, we have
used negativity and concurrence potential to compare the amount of
nonclassicality present in the linear and nonlinear QCS. 

In brief, a BS transformation can be described by
the Hamiltonian $H=\frac{1}{2}(a^{\dagger}b+ab^{\dagger})$, where
$a$ and $b$ are the annihilation operators of the two input modes.
Following Asboth's treatment, we combine an input state $\rho_{in}$
with the vacuum state $|0\rangle\langle0|$ at a symmetric BS
to obtain the output state as $\rho_{out}=U_{BS}\left(\rho_{in}\otimes|0\rangle\langle0|\right)U_{BS}^{\dagger}$,
where $U_{BS}=\exp[-iHt]$. 
Thus, the output state of the BS for the input state $|n\rangle\otimes|0\rangle$ can be obtained as \cite{ryl2017quantifying}
\begin{equation}
|n\rangle\otimes|0\rangle=|n,0\rangle\stackrel{{\rm BS}}{\longmapsto}\frac{1}{2^{n/2}}\sum_{j=0}^{n}\sqrt{^{n}C_{j}}\,\,|j,\,n-j\rangle.\label{eq:inout_fock}
\end{equation}
 Using Eq. (\ref{eq:inout_fock}) we can obtain the
output of the BS for a more general scenario where a finite superposition of Fock states described by  Eq. (\ref{eq:state1}) is inserted from one port of the BS while a vacuum state is inserted from the other port. In this case, the  two-mode output state is obtained as
\begin{equation}
|\psi\rangle_{d}\otimes|0\rangle=|\psi,0\rangle\stackrel{{\rm BS}}{\longmapsto}\sum_{n=0}^{d-1}\,\frac{c_{n}}{2^{n/2}}\sum_{j=0}^{n}\sqrt{^{n}C_{j}}\,\,|j,\,n-j\rangle.\label{eq:inout_psi}
\end{equation}
It is already stated (see Section \ref{sec:COMPARISON-OF-NONCLASSICALITY-WITNESSES})
that a finite superposition of Fock state is always nonclassical. Consequently we can expect
the output state (\ref{eq:inout_psi}) to be entangled \cite{asboth2005computable}.
Therefore, in what follows, we quantify the amount of entanglement
in the output state obtained in Eq. (\ref{eq:inout_psi}). 
\subsection{Negativity Potential }
\begin{figure}[h]
\centering
\subfigure[]{\includegraphics[scale=0.6]{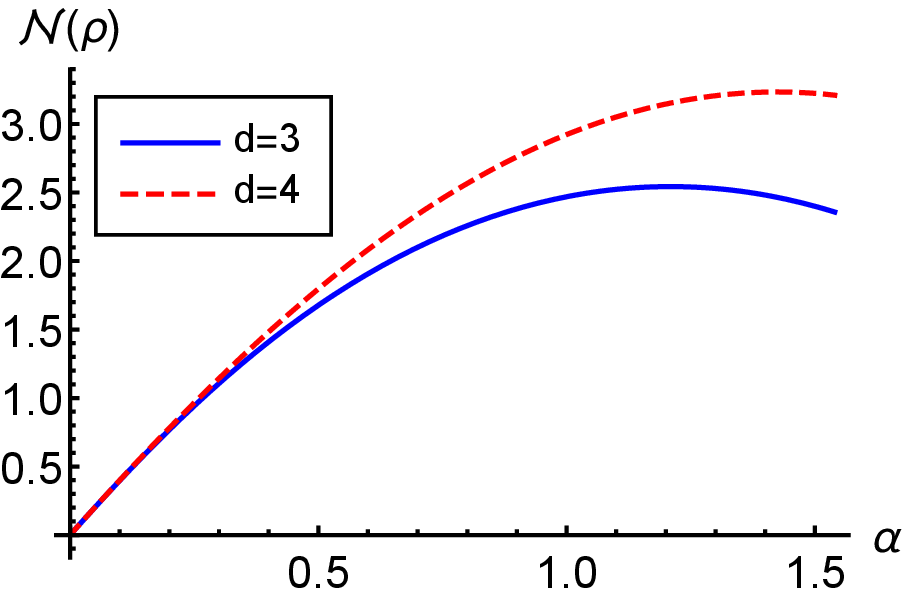}}\quad\quad
\subfigure[]{\includegraphics[scale=0.6]{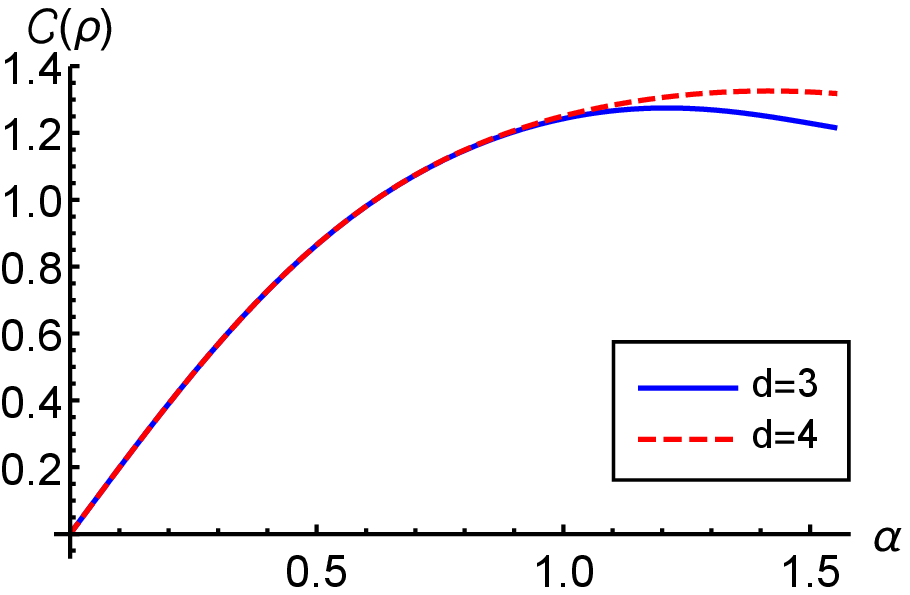}}\\
\subfigure[]{\includegraphics[scale=0.6]{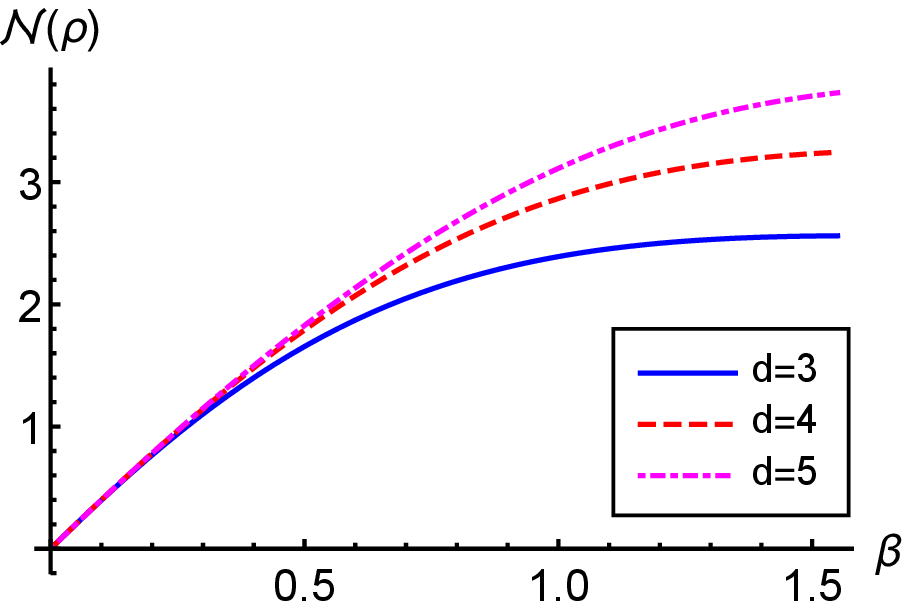}}\quad\quad
\subfigure[]{\includegraphics[scale=0.6]{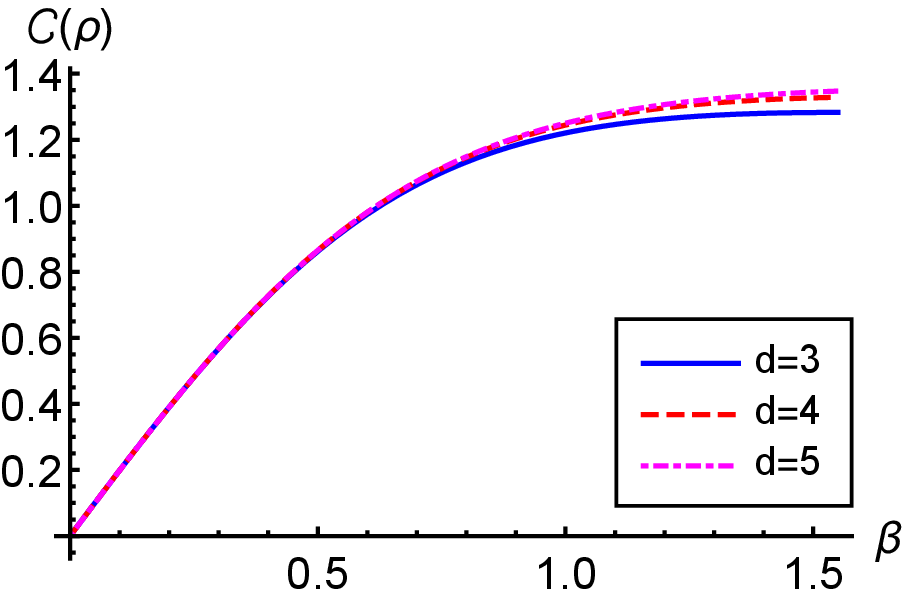}}
\caption{\label{fig:negativity_concurrence}The variation of negativity potential and
concurrence potential are shown with parameter $\alpha$ or $\beta$. (a)
and (c) show negativity potential in state $\left|\alpha\right\rangle _{d}$
and $\left|\beta\right\rangle _{d}$ for $d=3$ (smooth blue line) and
$d=4$ (dashed red line), respectively. (b) and (d) show concurrence in state $\left|\alpha\right\rangle _{d}$
and $\left|\beta\right\rangle _{d}$ for $d=3$ (smooth blue line),
$d=4$ (dashed red line) and $d=5$ (dotted dashed magenta line), respectively. }
\end{figure}

The negativity potential (${\rm E_{\mathcal{N}}}\left(\rho\right)$)
is one of the useful quantitative measure of nonclassicality, which
uses logarithmic negativity as the measure of entanglement. Specifically,
the negativity of the two mode entangled state with a density matrix
$\rho$ is defined as 
\[
\mathcal{N\left(\rho\right)}={\rm max}\left\{ 0,\,-2\,{\rm min}\,eig\left(\rho_{{\rm out}}^{\Gamma}\right)\right\} ,
\]
where $\rho_{{\rm out}}^{\Gamma}$ is the partial transpose of the
output state $\rho_{{\rm out}}$ in Eq. (\ref{eq:inout_psi}). Further,
logarithmic negativity, to quantify entanglement in the units of bits,
is defined as \cite{vidal2002computable} 
\begin{equation}
E_{\mathcal{N}}\left(\rho\right)=\log_{2}\left|\left|\rho_{out}^{\Gamma}\right|\right|_{1},\label{eq:logarithmic_negativity}
\end{equation}
where $\left|\left|\,\cdot\,\right|\right|_{1}$ is the trace norm.
It is related to negativity as $E_{\mathcal{N}}\left(\rho\right)=\log_{2}\left(2\mathcal{N}+1\right)$. Hence, using Eqs. (\ref{eq:inout_psi}) and (\ref{eq:logarithmic_negativity}) and doing a bit of algebra we can  obtain an analytic expression for the logarithmic negativity as 
\begin{equation}
E_{\mathcal{N}}\left(\rho\right)=2\log_{2}\sum_{n=0}^{d-1}\,\frac{\left|c_{n}\right|}{2^{n/2}}\left(\sum_{j=0}^{n}\sqrt{^{n}C_{j}}\right).\label{eq:log_negativity_expression}
\end{equation}
Using Eqs. (\ref{c_n}), (\ref{C_n_Beta}) and (\ref{eq:log_negativity_expression}) and simplifying, we obtain the analytic expression for the negativity potential. Further, we show the variation of logarithmic negativity with respect
to $\alpha$ and $\beta$ in Figs. \ref{fig:negativity_concurrence}(a)
and \ref{fig:negativity_concurrence}(c) where the positive values
of the negativity potential in both cases ensure that the output states
of the BS are entangled, and thus, in consistency with our expectation, both linear and nonlinear QCSs are found to be  nonclassical. We observed that with an increase in the dimensions
(i.e., for the larger value of Fock basis $|n\rangle$ superposition)
negativity potential for the both the cases increases logarithmically
and attain a maximum value. The present results are consistent with
that reported by Asboth for the Fock state \cite{asboth2005computable}.

\subsection{Concurrence potential}
Concurrence is a universal quantitative measure of entanglement \cite{wootters1998entanglement}.
For the finite-dimensional cases, the bipartite pure state $|\psi\rangle\in H_{A}\otimes H_{B}$
with ${\rm dim}\left[H_{A}\otimes H_{B}\right]<+\infty$, the concurrence
$\mathcal{C}(\rho)$ of $|\psi\rangle$ is defined as \cite{rungta2001universal}
\begin{equation}
\mathcal{C}(\rho)=\sqrt{2\left[1-{\rm Tr}\left(\rho_{A}^{2}\right)\right]},\label{eq:concurrence}
\end{equation}
where $\rho_{A}={\rm Tr_{B}}\left(|\psi\rangle\langle\psi|\right)$
is a mixed state which is obtained by taking the partial trace of
the output state of the BS. Therefore, in order to get
the concurrence of the finite superposition of Fock states, we need
to compute the value of $\rho_{A}^{2}$. The computation yields 
\[
{\rm Tr}\rho_{A}^{2}=\sum\limits _{n=0}^{d-1}\,\frac{\left|c_{n}\right|^{4}}{2^{2n}}\left\{ \sum\limits _{j=0}^{n}\left(^{n}C_{j}\right)^{2}\right\}, 
\]
and using this equation along with the Eqs. (\ref{c_n}), (\ref{C_n_Beta}), (\ref{eq:inout_psi}),
and (\ref{eq:concurrence}), we can obtain concurrence potential 
for the input nonclssical QCSs $|\alpha\rangle_{d}$
and $|\beta\rangle_{d}$.The obtained results are illustrated in Figs. \ref{fig:negativity_concurrence}(b)
and \ref{fig:negativity_concurrence}(d). 

\subsection{Anticlassicality\label{sec:Anticlassicality}}
\begin{figure}
\centering
\subfigure[]{\includegraphics[scale=0.4]{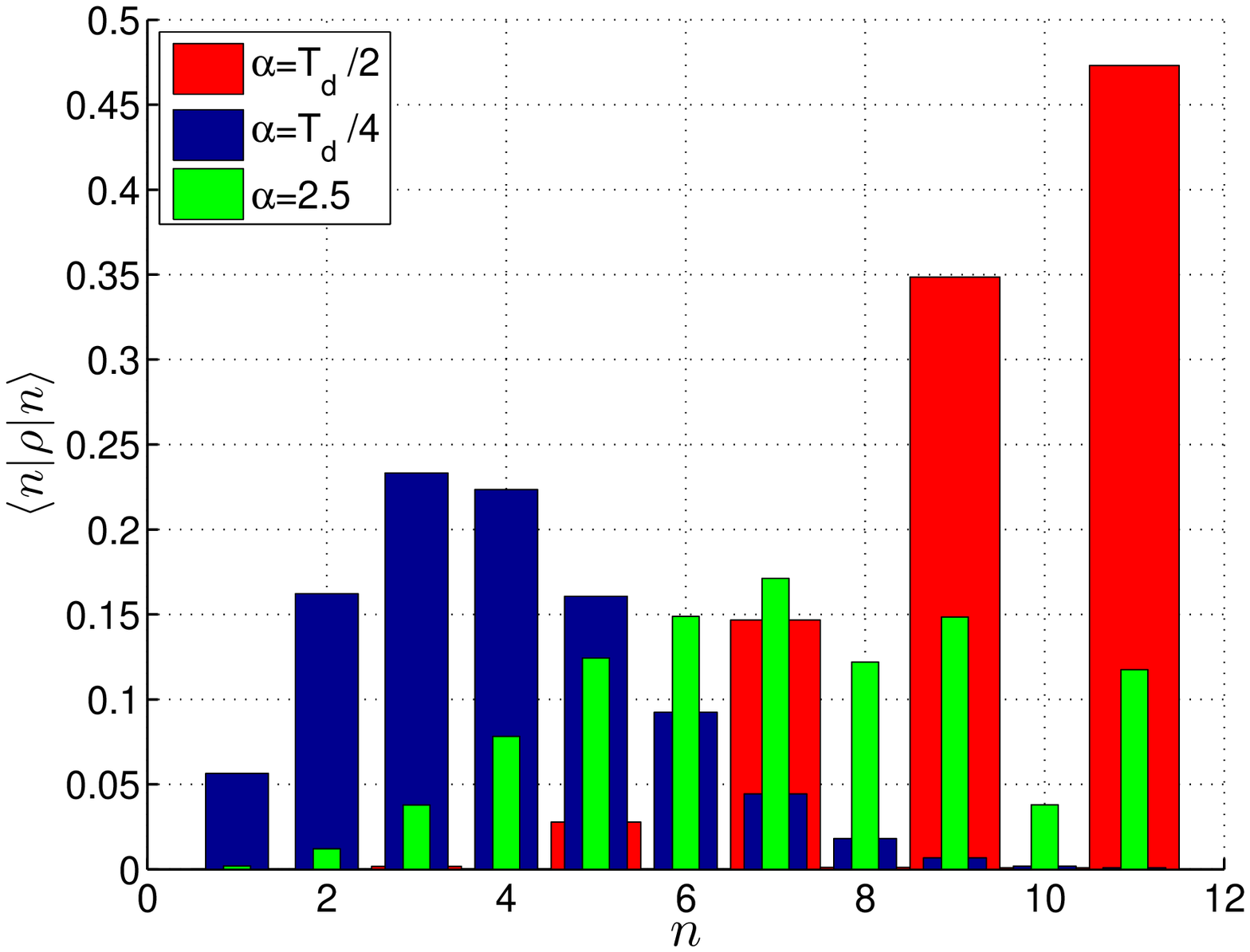}}\quad\quad
\subfigure[]{ \includegraphics[scale=0.4]{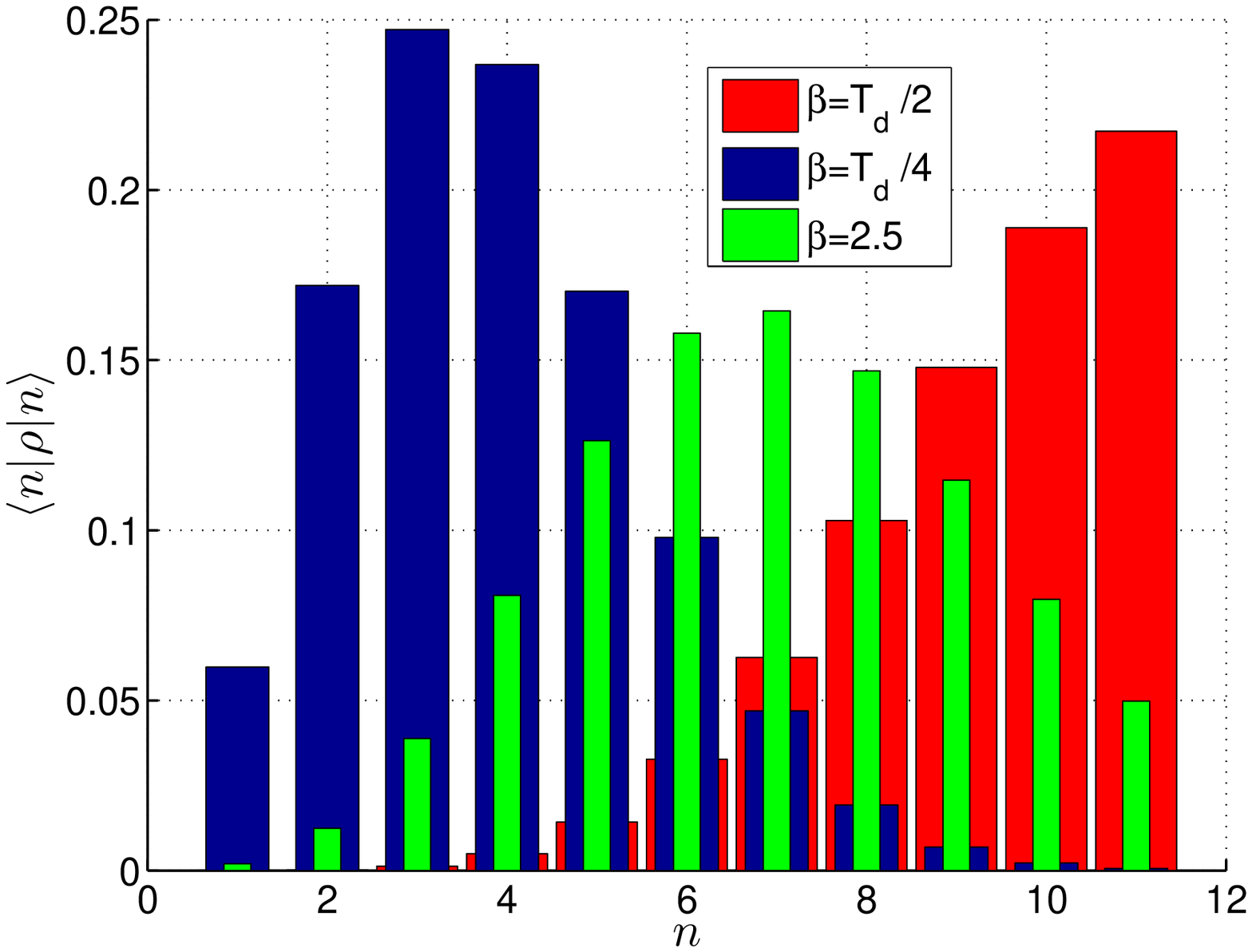}}\\
\caption{\label{fig:Anticlassiclaity}The variation of quantity $\langle n|\rho|n\rangle$
with respect to photon number  $(n)$, excluding $n=0$, where thick (red) bar, medium (blue) bar and thin (green) bar correspond
to $\beta=T_{d}/2, T_{d}/4$ and 2.5, respectively, for QCSs $|\alpha\rangle_{d}$
and $|\beta\rangle_{d}$ (in (a) and (b)). 
The highest values of the different set of bars $\left(\protect\underset{n}{{\rm max}}\langle n|\rho|n\rangle\right)$,
for each color, is the degree of anticlassiclaity $\left(\mathcal{A}_{1}\right)$}.
\end{figure}
Anticlassicality is a distance based measure of nonclassicality introduced by Dodonov
et al., in \cite{dodonov2003classicality}. It quantifies the amount of nonclassicality by measuring the distance
from the Fock states which are considered to be the most nonclassical states.  Now for any arbitrary
state $\rho$, the degree of anticlassicality is defined as \cite{dodonov2003classicality}
\begin{equation}
\mathcal{A}=\underset{n}{{\rm max}}\langle n|\rho|n\rangle,\label{eq:AntiClassicality}
\end{equation}
where the integer $n$ runs over all non-negative integers.
Excluding vacuum state $|0\rangle$, the degree of anticlassicality
is denoted as $\mathcal{A}_{1}$. In the present work, we have computed
the degree of anticlassicality for QCSs. The results are shown in
the Figs.\ref{fig:Anticlassiclaity}(a) and \ref{fig:Anticlassiclaity}(b),
where the maximum values of the bar correspond to the degree of the
anticlassicality $\left(\mathcal{A}_{1}\right)$ for linear and nonlinear QCSs. It is clear form the Figs. \ref{fig:Anticlassiclaity}(a) and \ref{fig:Anticlassiclaity}(b) that
for a given value of the $\alpha$ and $\beta$, 
$\mathcal{A}_{1}$s are different for QCSs. For the both QCSs, we observe that $\mathcal{A}_{1}$s are always  obtained at $n-1$ for $\alpha=\beta=T_{d}/2$.
The values of  $\mathcal{A}_{1}$s
for different values of $\alpha$ and $\beta$ for QCSs are calculated, and the results are  given in
the Table \ref{tab:Anticlassicalties--for}.

\section{CONCLUSION\label{sec:CONCLUSION}}
In conclusion, we would like to note that the present work reports the nonclassiclal features of linear and nonlinear QCSs through different witnesses of lower- and
higher-order nonclassicalties with a focus on the higher order nonclassicality. Further, it  quantifies the amount of nonclassicalities present in the linear and nonlinear QCSs by using negativity potential, concurrence potential, and anticlassicality. The uniqueness of this work lies in the fact that higher-order nonclassicalities of QCSs, had not been studied prior to the present work. Further, neither nonclassicalities present in QCS had been quantified earlier, nor anticlassicality had been used to quantify nonclassicality of similar systems in any of the existing works.

In the present work, various inequalities and measurement techniques are used which established the existence and quantification of the nonclassicality in QCSs. The obtained results are plotted and analyzed to reveal that in light of every nonclassical witness and measurement technique for linear and nonlinear QCSs have different characteristics.  It is well known
that any finite superposition of Fock states is nonclassical which
can be explained with the idea of the `hole burning' \cite{escher2004controlled}. However,
finite superposition of Fock states do not ensure the existence of
higher order nonclassicality. Here, we have explicitly established the existence of  higher-order nonclassicaliy in QCSs using various
witnesses  of higher- order nonclassicality, e.g., the criteria of HOA, HOS, HOSPS and Agarwal-Tara criterion.
In case of lower-order nonclassicality, we have used Klyshko's criterion and have compared the results (for both lower- and higher-order nonclassicalities) for linear and nonlinear
QCSs. 
 Interestingly, the comparison led to the observation that the nonclassicality witnesses show oscillation respect to $\alpha$ for nonlinear QCS, but no such oscillation is observed for linear QCS. 
We have also quantized the quantumness (amount of nonclassicality)
present in the QCSs by converting a nonclassical state to an entangled
state at the output of the BS. We calculate negativity potential and concurrence potential of the nonclassical state and we have shown that both of the cases amount of nonclassicality increases with the dimension ($d$) and for a particular dimension it gradually approaches a maximum value (cf. Figs. \ref{fig:negativity_concurrence}(a)-\ref{fig:negativity_concurrence}(d)). The anticlassicality for both types of QCSs are obtained and used for comparison (cf. Figs. \ref{fig:Anticlassiclaity}(a) and  \ref{fig:Anticlassiclaity}(b)). The degree of anticlassicality are also calculated for different values of $\alpha$ and $\beta$. The comparison result of the degree of anticlassicality for QCSs are given in the Table \ref{tab:Anticlassicalties--for}. The table (and also the results illustrated in Figs. \ref{fig:Anticlassiclaity}(a) and \ref{fig:Anticlassiclaity}(b)) restricts us from making a statement like linear (nonlinear) QCS is more nonclassical than the nonlinear (linear) QCS. The origin of this restriction on the comparative statement can be attributed to the fact that nonclassicality witnesses and measures are observed to oscillate only for nonlinear QCS.

 We conclude the paper with the hope that present work may be helpful for further research and the experimental findings specifically in the study of higher order nonclassicality for the QCSs having applications in the quantum information processing.

\begin{table}
\begin{centering}
\begin{tabular}{c|c|>{\centering}p{3.5cm}|>{\centering}p{3.5cm}}
\hline 
Serial number & Values of $\alpha$ or $\beta$ & Anticalssicality $(\mathcal{A}_{1})$ of  QCS $|\alpha\rangle$ & Anticalssicality $(\mathcal{A}_{1})$ of QCS $|\beta\rangle$\tabularnewline
\hline 
\hline 
$1.$ & $T_{d}/2$ & $0.473$ & $0.217$\tabularnewline
\hline
$2.$ & $T_{d}/4$ & $0.233$ & $0.247$\tabularnewline
\hline
$3.$ & $2.5$ & $0.171$ & $0.164$\tabularnewline
\hline 
\end{tabular}
\par\end{centering}
\caption{\label{tab:Anticlassicalties--for}Degree of anticlassicalty $\left(\mathcal{A}_{1}\right)$
for QCSs $|\alpha\rangle_{d}$ and $|\beta\rangle_{d}$ for different values
of  parameters $\alpha$ and $\beta$.}
\end{table}

\section*{Acknowledgment} A.P. and N.A. thank the Department of Science and Technology (DST),
India, for support provided through the DST project No. EMR/2015/000393.
A.P. also thanks K. Thapliyal and A. Miranowicz for some useful technical
discussions. 

\bibliographystyle{unsrt}
\bibliography{Qcs-final}

\begin{thebibliography}{10}

\bibitem{Agarwal2013quantum}
Girish~S. Agarwal.
\newblock {\em Quantum optics}.
\newblock Cambridge University Press, 2013.

\bibitem{klauder1985coherent}
John Klauder and B~Skagerstam.
\newblock {\em Coherent states: applications in physics and mathematical
  physics}.
\newblock World scientific, 1985.

\bibitem{zhang1990coherent}
Wei-Min Zhang, Robert Gilmore, et~al.
\newblock Coherent states: theory and some applications.
\newblock {\em Reviews of Modern Physics}, 62(4):867, 1990.

\bibitem{ficek2014quantum}
Zbigniew Ficek and Mohamed~Ridza Wahiddin.
\newblock {\em Quantum optics for beginners}.
\newblock CRC Press, 2014.

\bibitem{miranowicz2014phase}
Adam Miranowicz, Ma{\l}gorzata Paprzycka, Anirban Pathak, and Franco Nori.
\newblock Phase-space interference of states optically truncated by quantum
  scissors: Generation of distinct superpositions of qudit coherent states by
  displacement of vacuum.
\newblock {\em Physical Review A}, 89(3):033812, 2014.

\bibitem{miranowicz2014state}
Adam Miranowicz, Ji{\v{r}}{\'\i} Bajer, Ma{\l}gorzata Paprzycka, Yu-xi Liu,
  Alexandre~M Zagoskin, and Franco Nori.
\newblock State-dependent photon blockade via quantum-reservoir engineering.
\newblock {\em Physical Review A}, 90(3):033831, 2014.

\bibitem{sperling2014quantum}
J~Sperling, W~Vogel, and GS~Agarwal.
\newblock Quantum state engineering by click counting.
\newblock {\em Physical Review A}, 89(4):043829, 2014.

\bibitem{vogel1993quantum}
K~Vogel, VM~Akulin, and WP~Schleich.
\newblock Quantum state engineering of the radiation field.
\newblock {\em Physical Review Letters}, 71(12):1816, 1993.

\bibitem{miranowicz2004dissipation}
Adam Miranowicz and Wies{\l}aw Leo{\'n}ski.
\newblock Dissipation in systems of linear and nonlinear quantum scissors.
\newblock {\em Journal of Optics B: Quantum and Semiclassical Optics},
  6(3):S43, 2004.

\bibitem{pathak2013elements}
Anirban Pathak.
\newblock {\em Elements of quantum computation and quantum communication}.
\newblock Taylor \& Francis, 2013.

\bibitem{hillery2000quantum}
Mark Hillery.
\newblock Quantum cryptography with squeezed states.
\newblock {\em Physical Review A}, 61(2):022309, 2000.

\bibitem{alam2017lower}
Nasir Alam, Kishore Thapliyal, Anirban Pathak, Biswajit Sen, Amit Verma, and
  Swapan Mandal.
\newblock Lower-and higher-order nonclassicality in a {Bose}-condensed
  optomechanical-like system and a {Fabry-Perot} cavity with one movable
  mirror: squeezing, antibunching and entanglement.
\newblock {\em arXiv preprint arXiv:1708.03967}, 2017.

\bibitem{avenhaus2010accessing}
M~Avenhaus, K~Laiho, MV~Chekhova, and Ch~Silberhorn.
\newblock Accessing higher order correlations in quantum optical states by time
  multiplexing.
\newblock {\em Physical Review Letters}, 104(6):063602, 2010.

\bibitem{allevi2012measuring}
Alessia Allevi, Stefano Olivares, and Maria Bondani.
\newblock Measuring high-order photon-number correlations in experiments with
  multimode pulsed quantum states.
\newblock {\em Physical Review A}, 85(6):063835, 2012.

\bibitem{allevi2012high}
Alessia Allevi, Stefano Olivares, and Maria Bondani.
\newblock High-order photon-number correlations: a resource for
  characterization and applications of quantum states.
\newblock {\em International Journal of Quantum Information}, 10(08):1241003,
  2012.

\bibitem{pevrina2017higher}
Jan Pe{\v{r}}ina~Jr, V{\'a}clav Mich{\'a}lek, and Ond{\v{r}}ej Haderka.
\newblock Higher-order sub-poissonian-like nonclassical fields: Theoretical and
  experimental comparison.
\newblock {\em Physical Review A}, 96(3):033852, 2017.

\bibitem{abbott2016observation}
Benjamin~P Abbott, Richard Abbott, TD~Abbott, MR~Abernathy, Fausto Acernese,
  Kendall Ackley, Carl Adams, Thomas Adams, Paolo Addesso, RX~Adhikari, et~al.
\newblock Observation of gravitational waves from a binary black hole merger.
\newblock {\em Physical Review Letters}, 116(6):061102, 2016.

\bibitem{abbott2016gw151226}
BP~Abbott, R~Abbott, TD~Abbott, MR~Abernathy, F~Acernese, K~Ackley, C~Adams,
  T~Adams, P~Addesso, RX~Adhikari, et~al.
\newblock Gw151226: Observation of gravitational waves from a 22-solar-mass
  binary black hole coalescence.
\newblock {\em Physical Review Letters}, 116(24):241103, 2016.

\bibitem{harry2010advanced}
Gregory~M Harry, LIGO~Scientific Collaboration, et~al.
\newblock Advanced ligo: the next generation of gravitational wave detectors.
\newblock {\em Classical and Quantum Gravity}, 27(8):084006, 2010.

\bibitem{grote2013first}
H~Grote, K~Danzmann, KL~Dooley, R~Schnabel, J~Slutsky, and H~Vahlbruch.
\newblock First long-term application of squeezed states of light in a
  gravitational-wave observatory.
\newblock {\em Physical Review Letters}, 110(18):181101, 2013.

\bibitem{furusawa1998unconditional}
Akira Furusawa, Jens~Lykke S{\o}rensen, Samuel~L Braunstein, Christopher~A
  Fuchs, H~Jeff Kimble, and Eugene~S Polzik.
\newblock Unconditional quantum teleportation.
\newblock {\em Science}, 282(5389):706--709, 1998.

\bibitem{yuan2002electrically}
Zhiliang Yuan, Beata~E Kardynal, R~Mark Stevenson, Andrew~J Shields, Charlene~J
  Lobo, Ken Cooper, Neil~S Beattie, David~A Ritchie, and Michael Pepper.
\newblock Electrically driven single-photon source.
\newblock {\em science}, 295(5552):102--105, 2002.

\bibitem{bennett1993teleporting}
Charles~H Bennett, Gilles Brassard, Claude Cr{\'e}peau, Richard Jozsa, Asher
  Peres, and William~K Wootters.
\newblock Teleporting an unknown quantum state via dual classical and
  einstein-podolsky-rosen channels.
\newblock {\em Physical Review Letters}, 70(13):1895, 1993.

\bibitem{bennett1992communication}
Charles~H Bennett and Stephen~J Wiesner.
\newblock Communication via one-and two-particle operators on
  einstein-podolsky-rosen states.
\newblock {\em Physical Review Letters}, 69(20):2881, 1992.

\bibitem{ekert1991quantum}
Artur~K Ekert.
\newblock Quantum cryptography based on bell's theorem.
\newblock {\em Physical Review Letters}, 67(6):661, 1991.

\bibitem{miranowicz1994coherent}
A~Miranowicz, K~Piatek, and R~Tana{\'s}.
\newblock Coherent states in a finite-dimensional hilbert space.
\newblock {\em Physical Review A}, 50(4):3423, 1994.

\bibitem{leon1997finite}
W~Leonski.
\newblock Finite-dimensional coherent-state generation and quantum-optical
  nonlinear oscillator models.
\newblock {\em Physical Review A}, 55(5):3874, 1997.

\bibitem{buvzek1992coherent}
V~Bu{\v{z}}ek, AD~Wilson-Gordon, PL~Knight, and WK~Lai.
\newblock Coherent states in a finite-dimensional basis: Their phase properties
  and relationship to coherent states of light.
\newblock {\em Physical Review A}, 45(11):8079, 1992.

\bibitem{galetti1996discrete}
D~Galetti and MA~Marchiolli.
\newblock Discrete coherent states and probability distributions in
  finite-dimensional spaces.
\newblock {\em Annals of Physics}, 249(2):454--480, 1996.

\bibitem{kuang1993dynamics}
Le-Man Kuang, Fa-Bo Wang, and Yan-Guo Zhou.
\newblock Dynamics of a harmonic oscillator in a finite-dimensional hilbert
  space.
\newblock {\em Physics Letters A}, 183(1):1--8, 1993.

\bibitem{zhu1994even}
Jiu-Yun Zhu and Kuang Le-Man.
\newblock Even and odd coherent states of a harmonic oscillator in a
  finite-dimensional hilbert space and their squeezing properties.
\newblock {\em Physics Letters A}, 193(3):227--234, 1994.

\bibitem{roy1998coherent}
B~Roy and P~Roy.
\newblock Coherent states, even and odd coherent states in a finite-dimensional
  {Hilbert} space and their properties.
\newblock {\em Journal of Physics A: Mathematical and General}, 31(4):1307,
  1998.

\bibitem{miranowicz2003quantum1}
Adam Miranowicz, Wieslaw Leonski, and Nobuyuki Imoto.
\newblock Quantum-optical states in finite-dimensional hilbert space. i.
  general formalism.
\newblock {\em Modern Nonlinear Optics}, (Part I):155--193, 2003.

\bibitem{miranowicz2003quantum2}
Adam Miranowicz.
\newblock {Quantum-optical states in finite-dimensional Hilbert space. II.
  State generation}.
\newblock {\em Advances in Chemical Physics, Volume 119, Part 1: Modern
  Nonlinear Optics}, 155:195, 2003.

\bibitem{pegg1998optical}
David~T Pegg, Lee~S Phillips, and Stephen~M Barnett.
\newblock Optical state truncation by projection synthesis.
\newblock {\em Physical Review Letters}, 81(8):1604, 1998.

\bibitem{ozdemir2001quantum}
Sahin~Kaya Ozdemir, Adam Miranowicz, Masato Koashi, and Nobuyuki Imoto.
\newblock Quantum-scissors device for optical state truncation: A proposal for
  practical realization.
\newblock {\em Physical Review A}, 64(6):063818, 2001.

\bibitem{leonski2011quantum}
W~Leonski and A~Kowalewska-Kudlaszyk.
\newblock Quantum scissors--finite-dimensional states engineering.
\newblock {\em Progress in Optics}, 56:131, 2011.

\bibitem{miranowicz2013two}
Adam Miranowicz, Ma{\l}gorzata Paprzycka, Yu-xi Liu, Ji{\v{r}}{\'\i} Bajer, and
  Franco Nori.
\newblock Two-photon and three-photon blockades in driven nonlinear systems.
\newblock {\em Physical Review A}, 87(2):023809, 2013.

\bibitem{rabl2011photon}
Peter Rabl.
\newblock Photon blockade effect in optomechanical systems.
\newblock {\em Physical Review Letters}, 107(6):063601, 2011.

\bibitem{kuang1994coherent}
Le-Man Kuang, Fa-Bo Wang, and Yan-Guo Zhou.
\newblock Coherent states of a harmonic oscillator in a finite-dimensional
  hilbert space and their squeezing properties.
\newblock {\em Journal of Modern Optics}, 41(7):1307--1318, 1994.

\bibitem{miranowicz2015statistical}
Adam Miranowicz, Karol Bartkiewicz, Anirban Pathak, Jan Pe{\v{r}}ina~Jr,
  Yueh-Nan Chen, and Franco Nori.
\newblock Statistical mixtures of states can be more quantum than their
  superpositions: Comparison of nonclassicality measures for single-qubit
  states.
\newblock {\em Physical Review A}, 91(4):042309, 2015.

\bibitem{shchukin2005nonclassical}
Evgeny~V Shchukin and Werner Vogel.
\newblock Nonclassical moments and their measurement.
\newblock {\em Physical Review A}, 72(4):043808, 2005.

\bibitem{richter2002nonclassicality}
Th~Richter and W~Vogel.
\newblock Nonclassicality of quantum states: a hierarchy of observable
  conditions.
\newblock {\em Physical Review Letters}, 89(28):283601, 2002.

\bibitem{miranowicz2010testing}
Adam Miranowicz, Monika Bartkowiak, Xiaoguang Wang, Yu-xi Liu, and Franco Nori.
\newblock Testing nonclassicality in multimode fields: A unified derivation of
  classical inequalities.
\newblock {\em Physical Review A}, 82(1):013824, 2010.

\bibitem{lee1990many}
Ching~Tsung Lee.
\newblock Many-photon antibunching in generalized pair coherent states.
\newblock {\em Physical Review A}, 41(3):1569, 1990.

\bibitem{an2002multimode}
Nguyen~Ba An.
\newblock Multimode higher-order antibunching and squeezing in trio coherent
  states.
\newblock {\em Journal of Optics B: Quantum and Semiclassical Optics},
  4(3):222, 2002.

\bibitem{pathak2006control}
A~Pathak and ME~Garcia.
\newblock Control of higher order antibunching.
\newblock {\em Applied Physics B}, 84(3):479--484, 2006.

\bibitem{verma2010generalized}
Amit Verma and Anirban Pathak.
\newblock Generalized structure of higher order nonclassicality.
\newblock {\em Physics Letters A}, 374(8):1009--1020, 2010.

\bibitem{verma2008higher}
Amit Verma, Navneet~K Sharma, and Anirban Pathak.
\newblock Higher order antibunching in intermediate states.
\newblock {\em Physics Letters A}, 372(34):5542--5551, 2008.

\bibitem{thapliyal2017comparison}
Kishore Thapliyal, Nigam~Lahiri Samantray, J~Banerji, and Anirban Pathak.
\newblock Comparison of lower order and higher order nonclassicality in photon
  added and photon subtracted squeezed coherent states.
\newblock {\em Physics Letters A}, 381(37):3178 -- 3187, 2017.

\bibitem{thapliyal2014higher}
Kishore Thapliyal, Anirban Pathak, Biswajit Sen, and Jan Pe{\v{r}}ina.
\newblock Higher-order nonclassicalities in a codirectional nonlinear optical
  coupler: quantum entanglement, squeezing, and antibunching.
\newblock {\em Physical Review A}, 90(1):013808, 2014.

\bibitem{thapliyal2014nonclassical}
Kishore Thapliyal, Anirban Pathak, Biswajit Sen, and Jan Pe{\v{r}}ina.
\newblock Nonclassical properties of a contradirectional nonlinear optical
  coupler.
\newblock {\em Physics Letters A}, 378(46):3431--3440, 2014.

\bibitem{giri2014single}
Sandip~Kumar Giri, Biswajit Sen, C~H~Raymond Ooi, and Anirban Pathak.
\newblock Single-mode and intermodal higher-order nonclassicalities in two-mode
  {Bose-Einstein} condensates.
\newblock {\em Physical Review A}, 89(3):033628, 2014.

\bibitem{giri2017nonclassicality}
Sandip~Kumar Giri, Kishore Thapliyal, Biswajit Sen, and Anirban Pathak.
\newblock Nonclassicality in an atom--molecule {Bose--Einstein} condensate:
  Higher-order squeezing, antibunching and entanglement.
\newblock {\em Physica A: Statistical Mechanics and its Applications},
  466:140--152, 2017.

\bibitem{alam2015approximate}
Nasir Alam, Swapan Mandal, and Patrik {\"O}hberg.
\newblock Approximate analytical solutions of a pair of coupled anharmonic
  oscillators.
\newblock {\em Journal of Physics B: Atomic, Molecular and Optical Physics},
  48(4):045503, 2015.

\bibitem{alam2016nonclassical}
Nasir Alam and Swapan Mandal.
\newblock Nonclassical properties of coherent light in a pair of coupled
  anharmonic oscillators.
\newblock {\em Optics Communications}, 359:221--233, 2016.

\bibitem{alam2016quantum}
Nasir Alam and Swapan Mandal.
\newblock On the quantum phase fluctuations of coherent light in a chain of two
  anharmonic oscillators coupled through a linear one.
\newblock {\em Optics Communications}, 366:340--348, 2016.

\bibitem{sen2007quantum}
Biswajit Sen, Swapan Mandal, and Jan Pe{\v{r}}ina.
\newblock Quantum statistical properties of the radiation field in spontaneous
  {Raman} and stimulated {Raman} processes.
\newblock {\em Journal of Physics B: Atomic, Molecular and Optical Physics},
  40(7):1417, 2007.

\bibitem{sen2007squeezing}
Biswajit Sen and Swapan Mandal.
\newblock Squeezing, photon bunching, photon antibunching and nonclassical
  photon statistics in degenerate hyper raman processes.
\newblock {\em Journal of Physics B: Atomic, Molecular and Optical Physics},
  40(14):2901, 2007.

\bibitem{hillery1987amplitude}
Mark Hillery.
\newblock Amplitude-squared squeezing of the electromagnetic field.
\newblock {\em Physical Review A}, 36(8):3796, 1987.

\bibitem{hong1985higher}
CK~Hong and L~Mandel.
\newblock Higher-order squeezing of a quantum field.
\newblock {\em Physical Review Letters}, 54(4):323, 1985.

\bibitem{lee1990higher}
Ching~Tsung Lee.
\newblock Higher-order criteria for nonclassical effects in photon statistics.
\newblock {\em Physical Review A}, 41(3):1721, 1990.

\bibitem{agarwal1992nonclassical}
GS~Agarwal and K~Tara.
\newblock Nonclassical character of states exhibiting no squeezing or
  sub-poissonian statistics.
\newblock {\em Physical Review A}, 46(1):485, 1992.

\bibitem{klyshko1996nonclassical}
DN~Klyshko.
\newblock The nonclassical light.
\newblock {\em Physics-Uspekhi}, 39(6):573--596, 1996.

\bibitem{hillery1987nonclassical}
Mark Hillery.
\newblock Nonclassical distance in quantum optics.
\newblock {\em Physical Review A}, 35(2):725, 1987.

\bibitem{lee1991measure}
Ching~Tsung Lee.
\newblock Measure of the nonclassicality of nonclassical states.
\newblock {\em Physical Review A}, 44(5):R2775, 1991.

\bibitem{lutkenhaus1995nonclassical}
N~L{\"u}tkenhaus and Stephen~M Barnett.
\newblock Nonclassical effects in phase space.
\newblock {\em Physical Review A}, 51(4):3340, 1995.

\bibitem{kenfack2004negativity}
Anatole Kenfack and Karol {\.Z}yczkowski.
\newblock Negativity of the wigner function as an indicator of
  non-classicality.
\newblock {\em Journal of Optics B: Quantum and Semiclassical Optics},
  6(10):396, 2004.

\bibitem{asboth2005computable}
J{\'a}nos~K Asb{\'o}th, John Calsamiglia, and Helmut Ritsch.
\newblock Computable measure of nonclassicality for light.
\newblock {\em Physical Review Letters}, 94(17):173602, 2005.

\bibitem{vogel2014unified}
W~Vogel and J~Sperling.
\newblock Unified quantification of nonclassicality and entanglement.
\newblock {\em Physical Review A}, 89(5):052302, 2014.

\bibitem{ge2015conservation}
Wenchao Ge, Mehmet~Emre Tasgin, and M~Suhail Zubairy.
\newblock Conservation relation of nonclassicality and entanglement for
  {Gaussian} states in a beam splitter.
\newblock {\em Physical Review A}, 92(5):052328, 2015.

\bibitem{ryl2017quantifying}
S~Ryl, J~Sperling, and W~Vogel.
\newblock Quantifying nonclassicality by characteristic functions.
\newblock {\em Physical Review A}, 95(5):053825, 2017.

\bibitem{vidal2002computable}
Guifr{\'e} Vidal and Reinhard~F Werner.
\newblock Computable measure of entanglement.
\newblock {\em Physical Review A}, 65(3):032314, 2002.

\bibitem{wootters1998entanglement}
William~K Wootters.
\newblock Entanglement of formation of an arbitrary state of two qubits.
\newblock {\em Physical Review Letters}, 80(10):2245, 1998.

\bibitem{rungta2001universal}
Pranaw Rungta, Vladimir Bu{\v{z}}ek, Carlton~M Caves, Mark Hillery, and
  Gerard~J Milburn.
\newblock Universal state inversion and concurrence in arbitrary dimensions.
\newblock {\em Physical Review A}, 64(4):042315, 2001.

\bibitem{dodonov2003classicality}
VV~Dodonov and MB~Ren{\'o}.
\newblock Classicality and anticlassicality measures of pure and mixed quantum
  states.
\newblock {\em Physics Letters A}, 308(4):249--255, 2003.

\bibitem{escher2004controlled}
BM~Escher, AT~Avelar, TM~da~Rocha~Filho, and B~Baseia.
\newblock Controlled hole burning in the fock space via conditional
  measurements on beam splitters.
\newblock {\em Physical Review A}, 70(2):025801, 2004.

\end{thebibliography}
\end{document}